\documentclass[fleqn,usenatbib]{mnras}

\usepackage{newtxtext,newtxmath}

\usepackage[T1]{fontenc}

\DeclareRobustCommand{\VAN}[3]{#2}
\let\VANthebibliography\thebibliography
\def\thebibliography{\DeclareRobustCommand{\VAN}[3]{##3}\VANthebibliography}

\usepackage{graphicx}	
\usepackage{amsmath}	

\usepackage{amssymb}	
\usepackage{siunitx} 
\usepackage{booktabs}
\usepackage[flushleft]{threeparttable}
\usepackage{float}
\usepackage{cleveref}

\input{anc/ebcommands}
\newcommand{\PHOEBE}{\texttt{PHOEBE}}

\newcommand{\tsupconj}{$t_0$}
\newcommand{\teffratio}{$T_{\rm{eff},2}/T_{\rm{eff},1}$}
\newcommand{\requivsumfrac}{$\rho_1+\rho_2$}
\newcommand{\chiLIN}{$\chi^2_{1}$}
\newcommand{\chiFLAT}{$\chi^2_{0}$}
\newcommand{\RatioLT}{$\chi^2_{\rm{1}}/\chi^2_{\rm{0}}$}

\title[ASAS-SN Eclipsing Binaries]{The Value-Added Catalog of ASAS-SN Eclipsing Binaries:  Parameters of Thirty Thousand Detached Systems}

\author[D. M. Rowan et al.]{D. M.
Rowan,$^{1,2}$\thanks{E-mail: rowan.90@osu.edu},
T. Jayasinghe$^{1,2}$,
K. Z. Stanek$^{1,2}$,
C. S. Kochanek$^{1,2}$,
Todd A. Thompson$^{1,2,3}$,
\newauthor
B. J. Shappee$^{4}$,
T. W. -S. Holoien$^{5}$,
J. L. Prieto$^{6,7}$,
W. Giles$^{8}$
\\
$^{1}$Department of Astronomy, The Ohio State University, 140 West 18th Avenue, Columbus, OH, 43210, USA\\
$^{2}$Center for Cosmology and Astroparticle Physics, The Ohio State University, 191 W. Woodruff Avenue, Columbus, OH, 43210, USA\\
$^{3}$Department of Physics, The Ohio State University, Columbus, Ohio, 43210, USA\\
$^{4}$Institute for Astronomy, University of Hawaii, 2680 Woodlawn Drive, Honolulu, HI 96822, USA\\
$^{5}$Carnegie Observatories, 813 Santa Barbara Street, Pasadena, CA 91101, USA\\
$^{6}$N\'ucleo de Astronom\'ia de la Facultad de Ingenier\'ia y Ciencias, Universidad Diego Portales, Av. Ej\'ercito 441, Santiago, Chile\\
$^{7}$Millennium Institute of Astrophysics, Santiago, Chile\\
$^{8}$ASC Technology Services, 433 Mendenhall Laboratory 125 South Oval Mall Columbus OH, 43210, USA\\
}
\date{Accepted XXX. Received YYY; in original form ZZZ}

\pubyear{2022}

\begin{document}
\label{firstpage}
\pagerange{\pageref{firstpage}--\pageref{lastpage}}
\maketitle

\begin{abstract}
Detached eclipsing binaries are a fundamental tool for measuring the physical parameters of stars that are effectively evolving in isolation. Starting from more than 40,000 eclipsing binary candidates identified by the All-Sky Automated Survey for Supernovae (ASAS-SN), we use \PHOEBE{} to determine the sum of the fractional radii, the ratio of effective temperatures, the inclinations, and the eccentricities for \nSOL{} systems. We visually inspect all the light curve models to verify the model fits and examine the \textit{TESS} light curves, when available, to select systems with evidence for additional physics, such as spots, mass transfer, and hierarchical triples. We examine the distributions of the eclipsing binary model parameters and the orbital parameters. We identify two groups in the sum of the fractional radii and effective temperature ratio parameter space that may distinguish systems approaching the semidetached limit. Combining Gaia EDR3 with extinction estimates from 3-dimensional dust maps, we examine the properties of the systems as a function of their absolute magnitude and evolutionary state. Finally, we present light curves of selected eclipsing binaries that may be of interest for follow-up studies. 

\end{abstract}

\begin{keywords}
binaries:eclipsing -- surveys
\end{keywords}



\section{Introduction}

\begin{figure*}
    \centering
    \includegraphics[width=\linewidth]{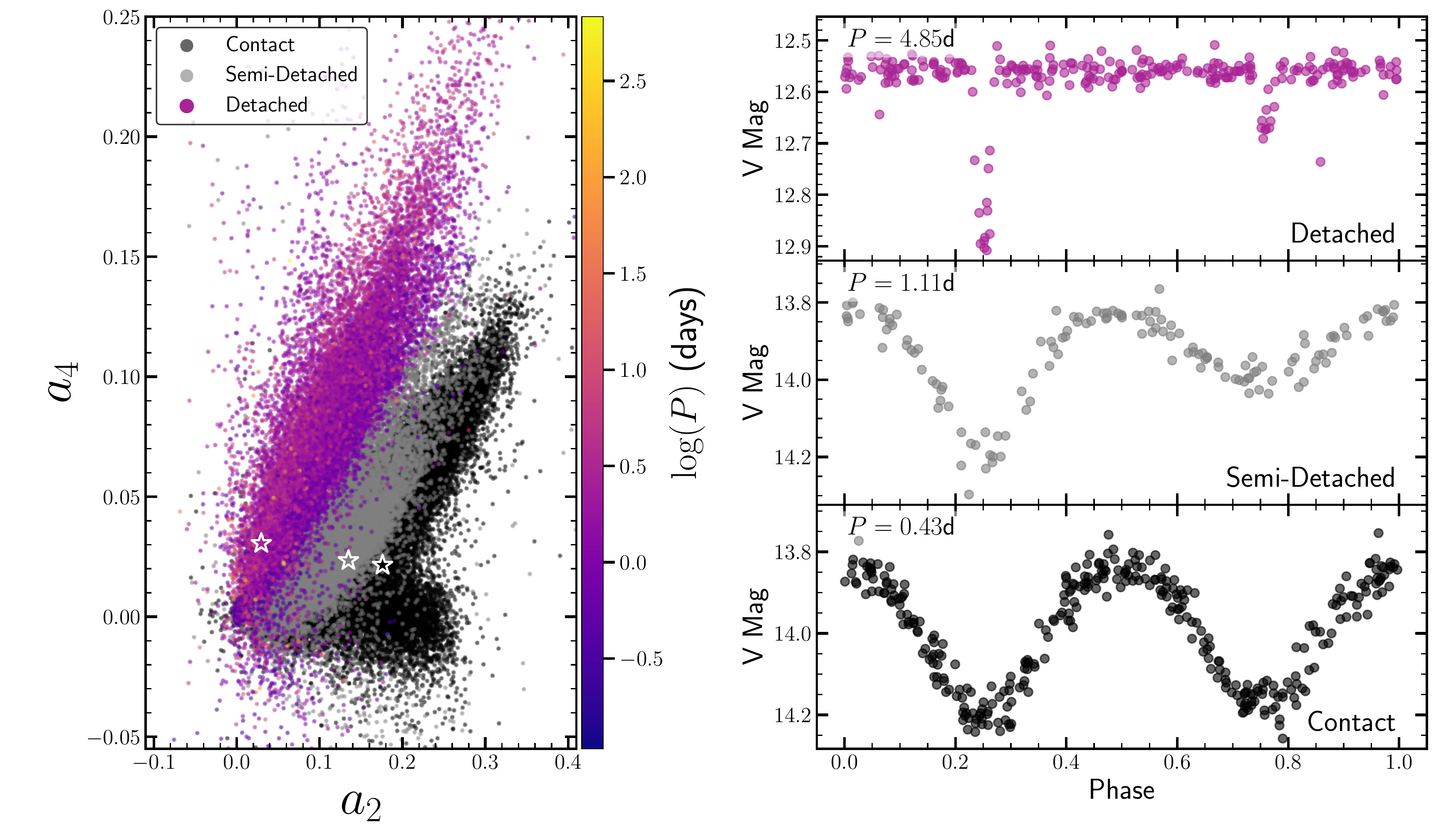}
    \caption{Fourier coefficients $a_2$ and $a_4$ for the eclipsing binaries from \citet{Jayasinghe21}. The detached eclipsing binaries included in our initial catalog are colored by period. We see that the shorter period binaries are distributed closer to the semi-detached population.  Examples of phase folded ASAS-SN $V$-band light curves are shown on the right and as stars on the left.}
    \label{fig:fourier_coeffs}
\end{figure*}

Detached eclipsing binaries (EBs) can be used to directly measure the stellar parameters of isolated stars. Modeling the light curve of an EB yields the sum of the fractional radii relative to the semimajor axis, $(R_1+R_2)/a$, the ratio of effective temperatures, the eccentricity, and the orbital inclination \citep{Kallrath09}. Physical masses and radii of the components can then be determined by adding radial velocity observations \citep[e.g. ][]{Matson17, Pourbaix04}. The resulting stellar parameters can be used to test and validate stellar evolution models \citep{Osterbrock53, Hoxie70, Andersen91, Pietrinferni04, Torres10, Feiden15} and as reddening and distance indicators \citep[e.g.,][]{Paczynski97, Wyithe02, Bonanos06, Pietrzynski09}.

Since most massive stars and nearly half of all Solar-type stars are found in binary systems \citep{Raghavan10, Sana12, Duchene13, Moe17}, binary evolution cannot be discounted when considering stellar evolution. During star formation, binarity impacts disk fragmentation and circumstellar disk formation \citep{Mathieu94, Tokovinin20}. Binary evolution on and after the main sequence can, often drastically, alter the evolution of the system. Tidal effects can circularize or synchronize the orbit \citep{Verbunt95, Hurley20} or induce oscillations in eccentric systems \citep{Kumar95, Willems02}. Mass transfer and common envelope evolution changes the orbital period, rotational periods, and luminosities of the components \citep{Paczynski76, Sana12, deMink13} and can also result in single stars through mergers \citep[e.g.,][]{Mateo90} or when one of the components explodes as a supernova \citep{Hoyle60}. Determining the physical parameters of binary systems at different stages of evolution is therefore crucially relevant in understanding how stellar populations evolve. 

The Optical Gravitational Lensing Experiment \citep[OGLE,][]{Graczyk11, Pawlak13, Pietrukowicz13, Soszynski16, Bodi21}, Kepler \citep{Prsa11, Slawson11, Kirk16}, the Wide-field Infrared Survey Explorer \citep[WISE,][]{Petrosky21}, and the All-Sky Automated Survey \citep[ASAS,][]{Pojmanski02, Paczynski06} have all produced extensive catalogs of EBs of varying morphologies and in different evolutionary stages. Selections from these catalogs have been used for more detailed photometric and spectroscopic modeling \citep[e.g.,][]{Ratajczak21}. 

Various codes exist to model the light curves of eclipsing binaries. \citep{Wilson71, Nelson72, Etzel81,Prsa05, Cokina21}. Although the physical radii cannot be determined without the inclusion of spectroscopic observations, the fractional radii can be measured from the eclipse widths. The ratio of effective temperatures is determined by the difference in eclipse depths, and the absolute temperatures can be determined with a multi-band light curve \citep{Prsa05, Torres10}. Light curve modeling tools have been applied to small selections of systems \citep[e.g.,][]{Popper81} and larger catalogs of EBs \citep[e.g.,][]{Devor05}. Construction of catalogs of models has aided in the study of binary stars of selected spectral types \citep[e.g.,][]{Bonanos04, Coughlin11, Graczyk18}, to select targets for radial velocity followup \citep[e.g.,][]{Matson17}, and to identify eccentric \citep{Bulut07, Kjurkchieva17, Kim18, Zasche21} and triple systems \citep{Hajdu22, Jurysek18}.

More than \num[group-separator={,}]{130000} eclipsing binaries have been identified in the All-Sky Automated Survey for Supernovae \citep[ASAS-SN,][]{Shappee14, Kochanek17} $V$-band data \citep{JayasingheII, Jayasinghe21}. Here we provide light curve models for the $g$- and $V$- band light curves of \nSOL{} detached EBs. Section \S\ref{sec:modeling} describes the ASAS-SN observations and the properties of the detached eclipsing binaries from \citet{JayasingheII} that form the basis of our catalog. Section \S\ref{sec:initial_filtering} describes the additional filtering applied to the $g$-band light curves and in Section \S\ref{sec:bls} we update the orbital periods. Section \S\ref{sec:rot} identifies systems with long-term trends. We use PHysics Of Eclipsing BinariEs \citep[\PHOEBE,][]{Prsa16, Conroy20} to model the ASAS-SN light curves in Section \S\ref{sec:models} and visually inspect the solutions in Section \S\ref{sec:visual_insepction}. Section \S\ref{sec:results} describes the statistics of the detached EB population and examines their empirical period-eccentricity distribution. In Section \S\ref{sec:period_distribution} we also explore their properties as a function of absolute magnitude and evolutionary state. 

\section{Modeling Detached Eclipsing Binaries} \label{sec:modeling}

\begin{figure*}
    \centering
    \includegraphics[width=0.6\linewidth]{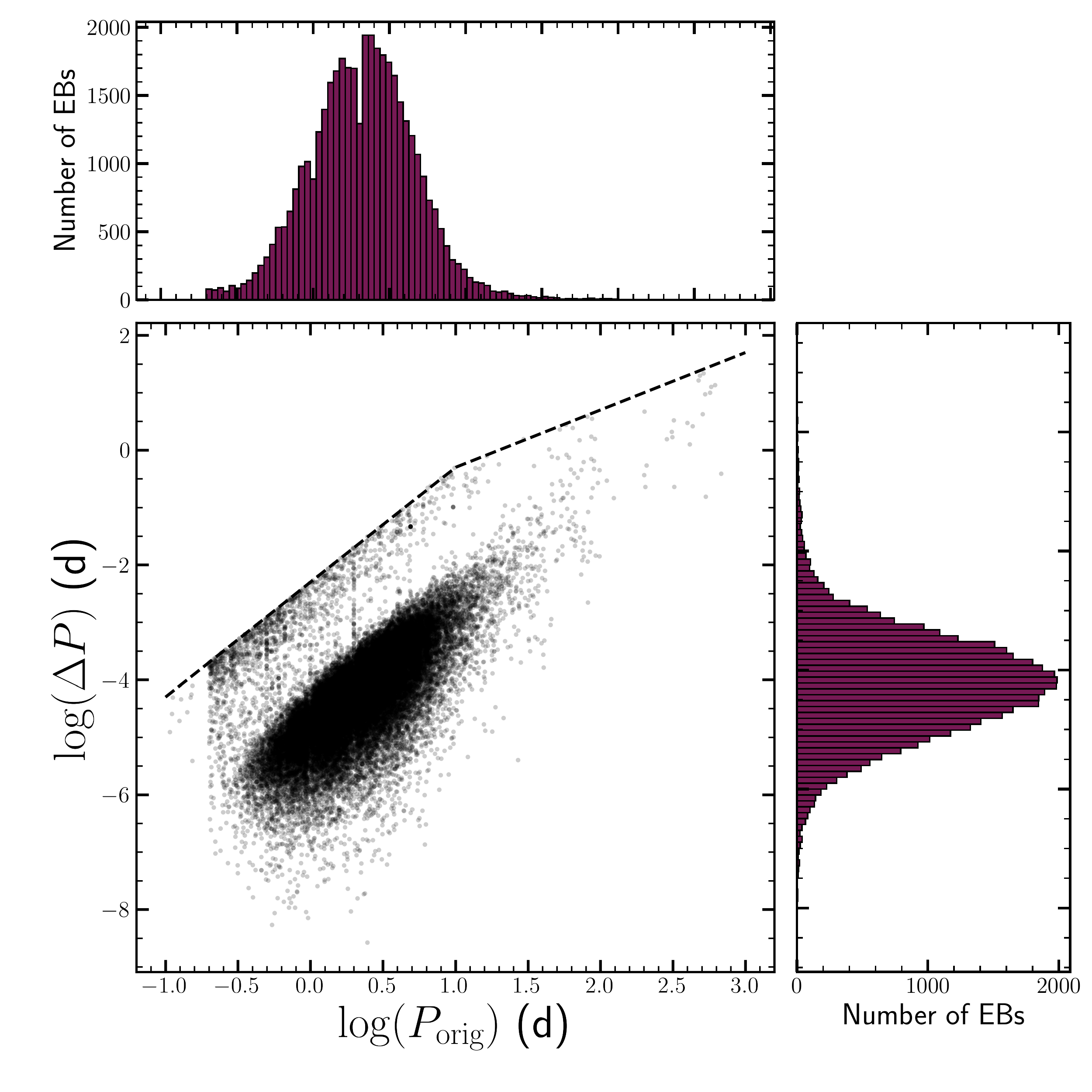}
    \caption{The distribution of period corrections $\Delta P$ as a function of the original period. The dashed line corresponds to $\Delta P_{\rm{max}}$. There is a small population of systems with $\Delta P \sim \Delta P_{\rm{max}}$ and when we visually inspect these light curves we find that the majority are poor quality EBs. The `notches' in the upper period histogram at 1 and 2~days are due to \citet{JayasingheII} filtering out these periods to minimize problems from diurnal aliasing.}
    \label{fig:bls}
\end{figure*}

The ASAS-SN $V$-band observations taken from 2012 to mid 2018 were used to classify more than 420,000 variable stars, including over $70,000$ contact binaries (EW, W Uma), $24,000$ semi-detached binaries (EB, $\beta$-Lyrae), and \nEA{} detached binaries (EA, Algol) \citep{JayasingheI, JayasingheII}. We select the detached population for our input catalog. At the end of 2017, ASAS-SN switched to observe in the $g$-band and expanded with three additional quadruple telescope units. We analyze both the $V$- and $g$-band light curves in this work. The ASAS-SN targets in our catalog have median $V$-band magnitudes ranging from \VmagMin{} to \VmagMax{}~mag with a median magnitude of \VmagMed{} mag. The optimal magnitude range for ASAS-SN targets is $11 < V < 17$~mag \citep{JayasingheII}. With the combined baseline of the $V$- and $g$-band observations, the EBs have a median observation baseline of \baselineMed{} years with an average of \epochsMean{} epochs. 

The three eclipsing binary morphology classes can be distinguished using the coefficients of a Fourier series model of their light curves \citep{Rucinski97, Paczynski06}. \citet{JayasingheII} included these coefficients in the random forest classification of ASAS-SN variable stars, and Figure \ref{fig:fourier_coeffs} shows the $a_2$ and $a_4$ coefficients for the three morphology classes with points colored by the orbital period for the detached systems. Unsurprisingly, shorter period systems classified as detached tend to be more similar to the semi-detached population. 


\subsection{Initial Data Filtering} \label{sec:initial_filtering}

Some ASAS-SN $g$-band light curves have additional problems from blends, detector edge effects, or saturation. The light curves of these targets can show spurious measurements with large errorbars, resulting in a bimodal magnitude error distribution. Although most of the affected data points can be removed by sigma clipping, we screen these targets before we attempt to model their light curves. 

To select targets with biomdal error distributions, we search for stars with broad distributions of magnitude errors. In general, fainter targets and EBs with deep eclipses will also have broad magnitude error distributions, so we visually inspect targets that have large standard deviations in their magnitude error distributions. In total, we flag \nMEC{} targets for additional filtering based on the magnitude error distribution. For these targets, we use {\tt sklearn} \citep{Pedregosa11} to fit a two-component Gaussian mixture model to the magnitude error distribution. We then remove observations that have a probability $p>0.95$ of belonging to the higher error component. 

\subsection{Orbital Period Calculation} \label{sec:bls}

\begin{figure*}
    \centering
    \includegraphics[width=0.6\linewidth]{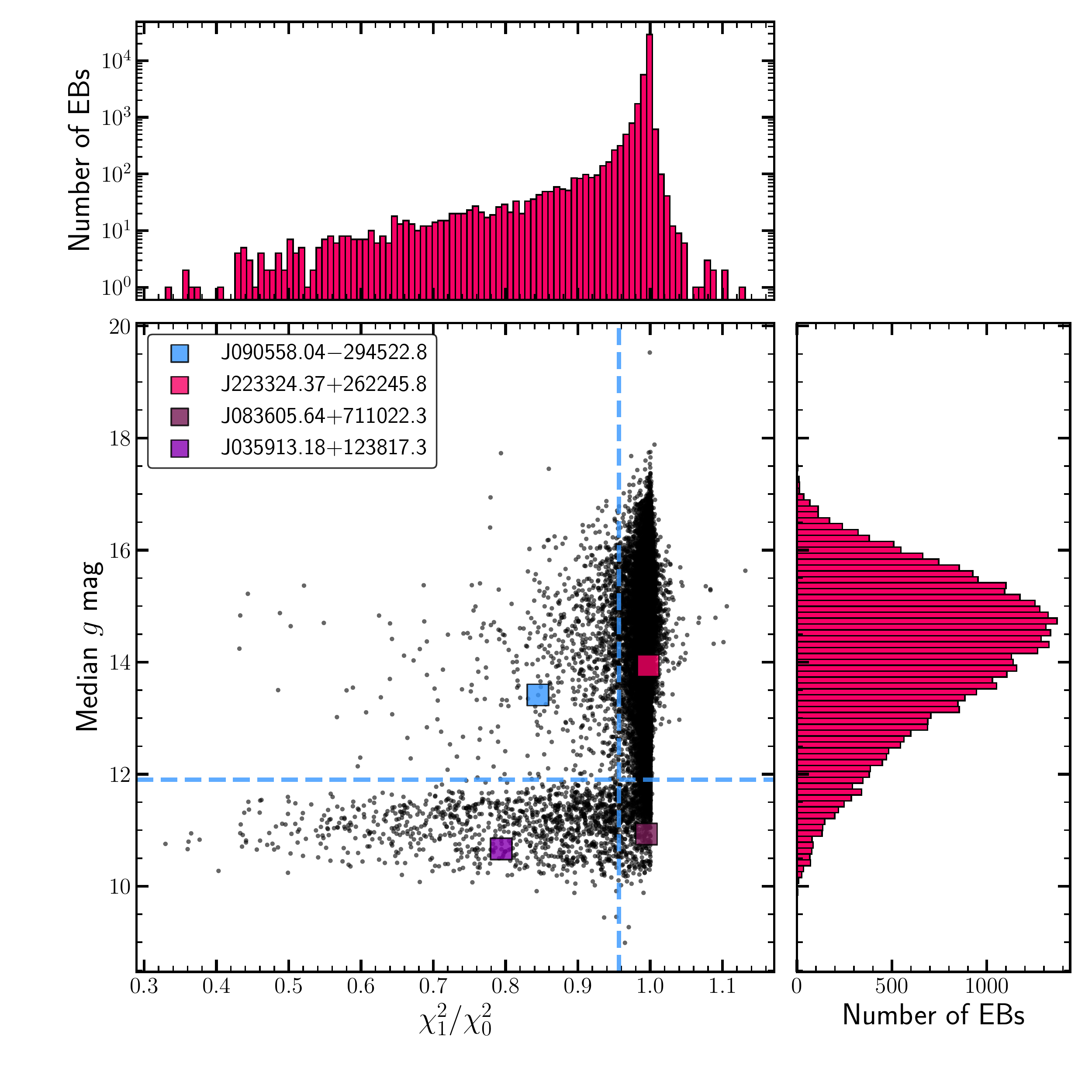}
    \caption{Long-term trends in the ASAS-SN $g$-band light curves may be evidence of rotational variability from spots on one or both stars. We also find that bright $g$-band targets are more likely to have long-term trends because they are near saturation of the CCDs. Top: ratio \RatioLT{} for our EA catalog. Right: median $g$-band magnitude. Center: for targets to the left of the vertical line we modeled the $V$-band light curve. Targets in the upper left quadrant are flagged as potential rotational variables in our final catalog. We mark four examples and show the unfolded light curves in Figure \ref{fig:eb_rot_examples}.}
    \label{fig:eb_rot_scatter}
\end{figure*}

\citet{JayasingheII} determined the orbital periods of the detached EBs with a combination of the Generalized Lomb Scargle \citep{Lomb76, Scargle82}, Multi-Harmonic Analysis of Variance \citep{SchwarzenbergCzerny96}, and Box Least Squares (BLS) periodograms \citep{Kovacs02}. In order to correctly phase the data over the much longer time baseline of the $V$- and $g$-band data, these periods need to be updated. We use the {\tt astrobase} implementation of the BLS algorithm \citep{Bhatti18, Kovacs02} to search a narrow period window centered on the $V$-band catalog period, $P_{\rm{orig}}$, of width $2\Delta P$, where $\Delta P=0.005P_{\rm{orig}}^2$. We chose this narrow window for computational speed and increase the width as $P_{\rm{orig}}^2$ since the period errors will be larger for longer period EBs. To prevent returning a period that is a small integer ratio of the input period, we further restrict the search range to $0.95 P_{\rm{orig}} < P < 1.05P_{\rm{orig}}$, which only affects systems with $P_{\rm{orig}}>10$~d. 

Before computing the period, we first median normalize the fluxes in each band and use sigma clipping to remove observations fainter than (brighter than) 8.0 (2.0) standard deviations from the median magnitude. The $V$-band and $g$-band data are then combined and the BLS is run with 200 phase bins and eclipse durations ranging from 0.005 to 0.200 in orbital phase units. We select the peak with the highest power as the updated orbital period.

We also use the {\tt astropy} implementation of the \citep{Kovacs02} BLS \citep{Astropy13} to double check the periods. While the two implementations generally return periods matching to within $\sim10^{-4}-10^{-5}$~days, we find some cases where the {\tt astrobase} KBLS implementation is unable to identify the correct period and we instead use the {\tt astropy} period. Figure \ref{fig:bls} shows the distributions of $P_{\rm{new}}$ and $\Delta P$. The EBs span a wide range of periods from \minP{}~d to \maxP{}~d with a median of \medP{}~d. There are fewer systems with periods near 1 and 2 days because \citet{JayasingheII} dropped systems with these periods to minimize problems from diurnal aliasing. 

Some EBs have a much larger $\Delta P$ than expected, forming the small population of objects near $\Delta P_{\rm{max}}$ in Figure \ref{fig:bls}. We visually inspect all EA light curves in Section \S\ref{sec:visual_insepction}, and find that many of these systems are misclassified as EAs, or have poor $g$-band light curves because they are bright and close to saturation ($g < 11$~mag). While the ASAS-SN image pipeline does attempt to correct for saturation, the procedure does not always work well \citep[see][]{Kochanek17}.

\subsection{Long-term trends in EB light curves} \label{sec:rot}

\begin{figure*}
    \centering
    \includegraphics[width=\linewidth]{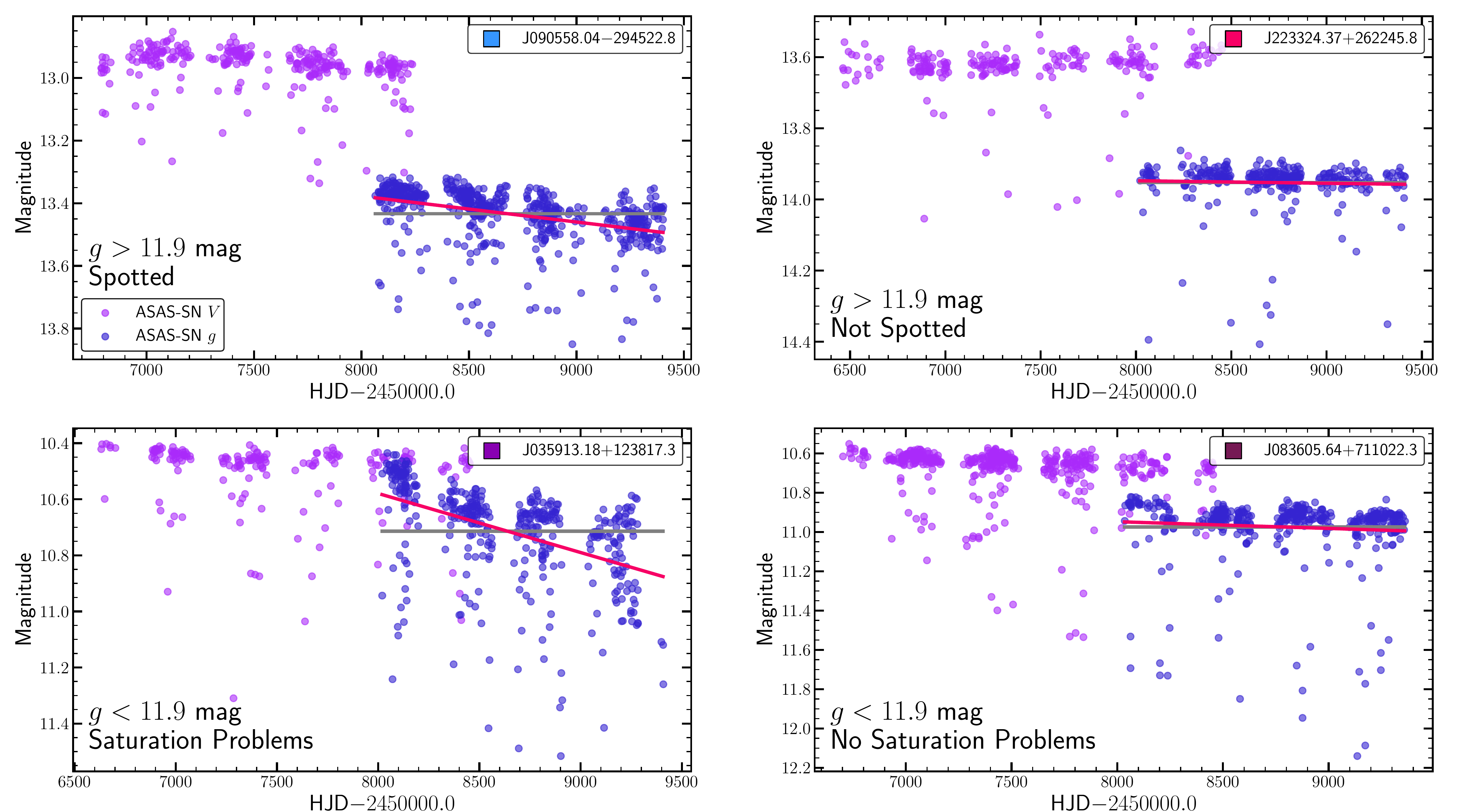}
    \caption{Example light curves for the four quadrants of Figure \ref{fig:eb_rot_scatter}. The gray and red lines show the flat line and linear fits, respectively. The panels are arranged to follow the quadrants of Figure \ref{fig:eb_rot_scatter} and are labeled by their likely properties.}
    \label{fig:eb_rot_examples}
\end{figure*}

Rotational variability can result in long-period trends in ASAS-SN $g$-band light curves from spot modulation \citep[e.g.,][]{Rowan21, Christy21}. Since the filter is bluer and the calcium H and K lines, with rest wavelengths 3969~\AA~and~3934~\AA, respectively, are associated with chromospheric activity and lie in the $g$-band (3858--5686~\AA), rotational modulations are usually more prominent in the $g$-band than in the $V$-band. The initial \PHOEBE{} light curve modeling steps are done on the $g$-band data before adding the $V$-band light curve for simultaneous optimization (Section \S\ref{sec:models}). For $g$-band light curves with evidence of rotational variability, we instead use the $V$-band light curve for the initial modeling steps since these light curves have less `noise'.

Following \citet{Rowan21}, we use a $\chi^2$ ratio test to identify $g$-band light curves with long term trends as an indication of rotational variability. We compute the \chiFLAT{} of a flat line and the \chiLIN{} of a linear fit in time for each light curve. EBs with some rotational variability tend to have  $R_{\rm{LT}}=$\RatioLT{}$<1$. Figure \ref{fig:eb_rot_scatter} shows the distribution of the systems in $R_{\rm{LT}}$ for the $g$-band light curves. Many of the EBs with long term trends are in the bright tail of the magnitude distribution, suggesting that stars approaching saturation also show systematic effects that produce long-term trends in the $g$-band light curves. We start the \PHOEBE{} modeling with the $V$-band light curves if $R_{\rm{LT}}<\cutROT{}$, corresponding to the bottom 5\% of the distribution (\nVbase{} systems). We also use the boundary of 11.9~mag to separate spotted systems from saturation problems. The \nROT{} EBs in the upper left quadrant defined by these two cuts are flagged as having potential rotational variability in Table \ref{tab:ea_table}.

Figure \ref{fig:eb_rot_examples} shows examples of the $V$- and $g$-band light curves for each quadrant. These four EBs are also marked in the main panel of Figure \ref{fig:eb_rot_scatter}. In some cases, such as ASASSN-V J090558.04$-$294522.8, long-term variation is also seen in the $V$-band data, further suggesting its astrophysical nature. On the other hand, ASASSN-V~J035913.18$+$123817.3 (lower left) has a $g$-band magnitude that decreases in time, but lies in the saturated region ($g=\qIVgmag{}$~mag) and there is no such variability seen in the $V$-band data, suggesting that the $g$-band trend is due to systematic effects.

\subsection{Eclipsing Binary Models} \label{sec:models}

We use \PHOEBE{} \citep{Prsa05, Prsa16, Conroy20} to model the ASAS-SN light curves. \PHOEBE{} has been used extensively for modeling contact binaries \citep[e.g.,][]{Kobulnicky22}, detached systems \citep[e.g.,][]{Way21}, and exotica such as heartbeat systems \citep[e.g.,][]{Ou21}.

We start by constructing a phased, flux normalized light curve and sigma clip points $>5\sigma$ from the median light curve. We use the geometry estimator in \PHOEBE{} to estimate the eclipse locations, and expand the limit to $8\sigma$ during the eclipses to prevent unwanted clipping of narrow, deep eclipses. During visual inspection (Section \ref{sec:visual_insepction}) we check for systems where over-clipping erroneously removes points during eclipses and leads to inaccurate parameter determination. The geometry estimator combines a two Gaussian model with a cosine term to fit for the eccentricty, $e$, argument of periastron, $\omega$, and time of superior conjunction, \tsupconj{} \citep{Mowlavi17, Conroy20}. 

We do the geometry estimator fits for periods of $P_{\rm{new}}/2$, $P_{\rm{new}}$, and $2P_{\rm{new}}$ because if the secondary eclipse is very shallow the BLS period can be off by a factor of two. If the ratio of the minimum $\chi^2$ of the fits to the second smallest $\chi^2$ is less than 0.8, we accept the period corresponding to the minimum to be the period $P$. Otherwise, we optimize the two best periods and select the fit with the minimum $\chi^2$. This additional step is effective at determining the correct period for systems with large eclipse depth ratios where the two-Gaussian model would otherwise return a satisfactory fit at $2P_{\rm{new}}$.

We also apply the EBAI estimator \citep{Prsa08} to the median filtered light curve to obtain initial values for the ratio of effective temperatures, \teffratio, the sum of fractional radii, $R_1/a + R_2/a=\rho_1+\rho_2$, and the inclination, $i$. We use the results from the geometry estimator to define the orbital phase range as $-0.5$ to $0.5$ with the primary eclipse at phase zero. We then use the {\tt polyfit} algorithm \citep{Prsa08} to transform the light curve into 200 points equally spaced in phase. After running both estimators, we have the initial values of $e$, \tsupconj, $\omega$, \teffratio, \requivsumfrac, and $i$ for a full optimization. 

Next we optimize the model of the $g$-band light curve ($V$-band for systems with long-term trends identified in Section \S\ref{sec:rot}) with 400 iterations of the Nelder-Mead simplex algorithm \citep{Gao12}. This is typically sufficient to converge to a minimum for the ASAS-SN light curves. We use the {\tt ellc} backend \citep{Maxted16} with no irradiation. The limb darkening coefficients are from the Castelli-Kurucz model atmosphere tables \citep{Castelli03} and span $3500 < T_{\rm{eff}} < 50000$~K. Since only the relative temperatures, rather than the absolute temperatures, are constrained by a single-band light curve fit, we solve for the \teffratio{} and keep $T_{\rm{eff},1}=7000$~K fixed to allow a wide range of values for $T_{\rm{eff},2}$ within the constraints of the Castelli–Kurucz model atmospheres. We then add in the $V$- or $g$-band data and run an additional 200 iterations on the multi-band light curve. The optimization with the additional light curve further constrains the model parameters and is especially effective at improving the fits for systems with narrow eclipses.

\subsection{Visual Inspection} \label{sec:visual_insepction}

\begin{figure}
    \centering
    \includegraphics[width=\linewidth]{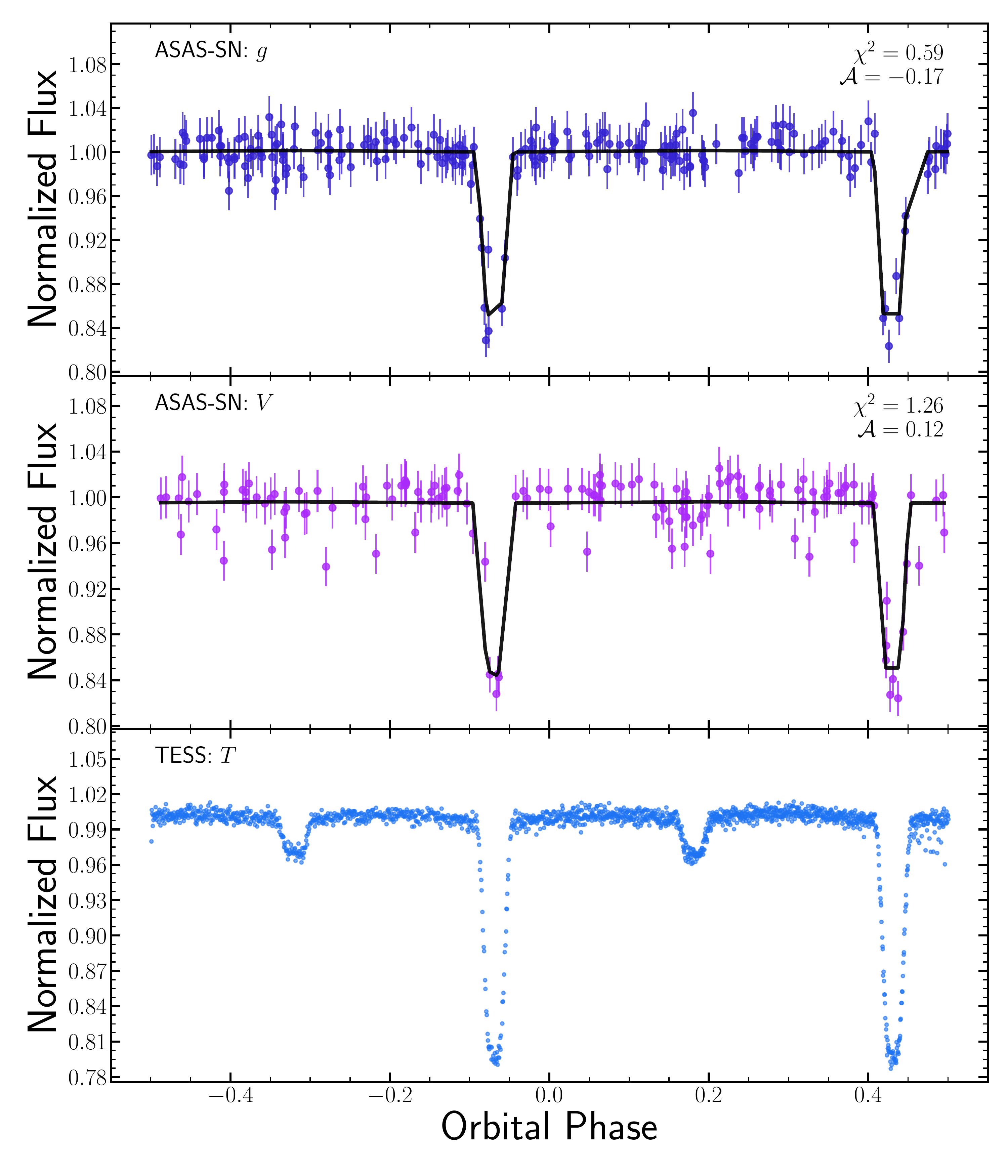}
    \caption{ASAS-SN $V$, $g$, and TESS $T$ band light curves for ASASSN-V J010018.84+552507.1. The period selection procedure described in Section \S\ref{sec:models} returned a period of 4.303092 days, but after visual inspection of the three light curves it is clear the correct period is instead 2.151546 days. The reduced $\chi^2_\nu$ and alarm statistic $\mathcal{A}$ are given for the $V$- and $g$-band fits.}
    \label{fig:ea_half_p}
\end{figure}

\begin{figure}
    \centering
    \includegraphics[width=\linewidth]{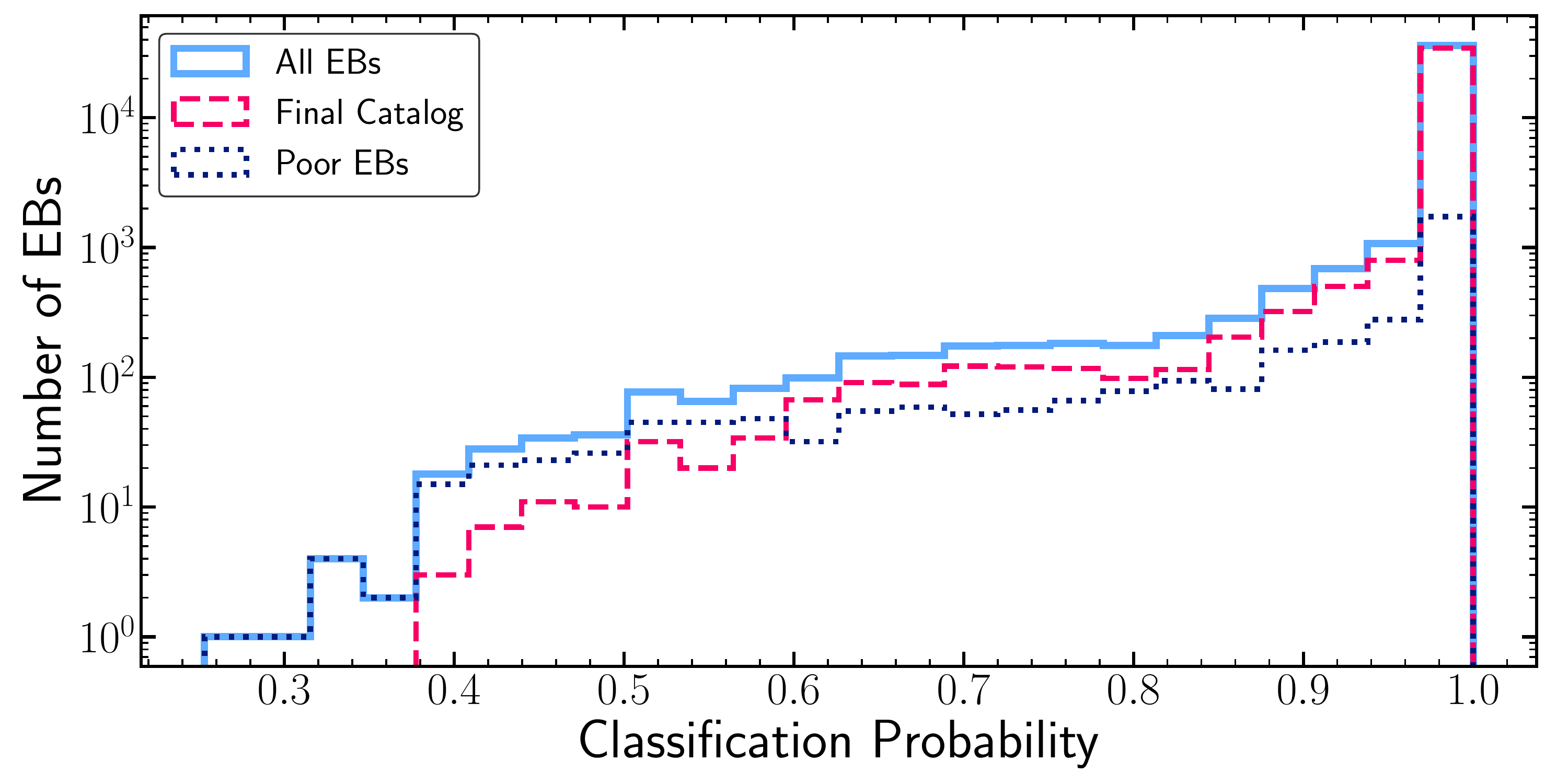}
    \caption{Distribution of EB classification probabilities from \citep{JayasingheII}. Out of the \nPOOR{} systems we flag as poor EBs, \LowCPPoor{} have $P_{\rm{class}}<0.9$.}
    \label{fig:class_probability} 
\end{figure}

\begin{table*}
    \centering
    \caption{Examples of the detached eclipsing binary model fit results sorted by orbital period. The time of superior conjunction ($t_0$), orbital period ($P$), sum of the fractional radii (\requivsumfrac{}), ratio of effective temperatures (\teffratio{}), eccentricity ($e$), argument of periastron ($\omega$), and inclination ($i$), are from the \PHOEBE{} models. The Gaia EDR3 color and absolute magnitude are corrected for extinction using {\tt mwdust} \citep{Bovy16}. The reduced $\chi^2_{\nu}$ is given for the optimized solution. The alarm statistics $\mathcal{A}_V$ and $\mathcal{A}_g$ (Equation \ref{eqn:alarm}) are given for the $V$- and $g$-band fits. Systems with long-term trends that are likely due to rotational variability are marked in the ROT column. Finally, we use the MIST isochrones and evolutionary tracks \citep{Choi16, Dotter16} to classify the systems based on the evolutionary state of the photometric primary (MS=main sequence, SG=subgiant, RG=giant). The full catalog is available as an online table.}
    \label{tab:ea_table}
    \sisetup{table-auto-round,
             group-digits=false}
    \setlength{\tabcolsep}{2pt}
    \begin{center}
    \begin{tabular}{l S[table-format=3.5] S[table-format=3.5] S[table-format=1.2] S[table-format=1.2] S[table-format=1.3] S[table-format=3.2] S[table-format=2.1] S[table-format=1.2] S[table-format=2.2] S[table-format=2.2] S[table-format=2.2] S[table-format=2.2] S[table-format=2.1] r r}
\toprule
     {ASAS-SN Name} &     {$t_0$} &  {Period} &  {$\rho_1+\rho_2$} &  {$T_{\rm{eff},2}/T_{\rm{eff},1}$} &  {Ecc} &  {$\omega$} &    {Incl} &   {$M_G$} & {$M_{\rm{BP}}-M_{\rm{RP}}$} &  {$\chi^2_{\nu}$} &  {$\mathcal{A}_V$} &  {$\mathcal{A}_g$} &  {$g$} &       {ROT?} & {State} \\
{} & {$+2456000$ (d)} & {(d)} & {} & {} & {} & {(deg)} & {(deg)} & {(mag)} & {(mag)} & {} & {} & {} & {(mag)} & {} & {}\\ 
\midrule
J063849.93-090909.2 & 2333.590990 & 11.940168 &           0.238009 &                           0.938228 &  0.421 &  140.196615 & 83.865412 & -0.442985 &                    0.260126 &          1.246927 &           1.000634 &           3.612254 & 14.060 &              &         \\
J000538.47+360212.4 & 2024.275445 &  9.558278 &           0.150087 &                           0.999202 &  0.013 &   88.294888 & 88.160127 &  2.006321 &                    0.612154 &          0.900750 &           6.460094 &           1.282864 & 12.611 &              &      MS \\
J184413.44-372828.3 & 2016.346875 &  9.283950 &           0.291023 &                           0.528277 &  0.010 &   78.027319 & 83.784214 &  1.747878 &                    0.773098 &          1.070620 &           3.440331 &           1.004776 & 13.026 &              &      SG \\
J174224.72+220501.8 & 2037.687606 &  7.359367 &           0.174317 &                           0.975428 &  0.016 &   27.849991 & 87.394418 &   2.86254 &                    0.834785 &          1.054890 &           0.365046 &           0.313307 & 13.773 &              &      SG \\
J054047.45-692028.9 & 2039.965990 &  5.413941 &           0.555445 &                           0.946371 &  0.031 &  288.901211 & 76.807926 &  -3.38331 &                   -2.509675 &          0.749716 &          -0.079716 &          -0.061100 & 13.774 &              &         \\
J224958.20-473207.0 & 2019.961077 &  5.047867 &           0.305650 &                           0.725314 &  0.026 &  259.317510 & 85.408962 &  3.239277 &                    1.118726 &          2.667935 &           2.599394 &           0.698147 & 14.552 &              &      SG \\
J150133.03-442523.4 & 1464.784148 &  4.655767 &           0.264959 &                           0.900866 &  0.016 &  279.917141 & 84.783619 &  2.622998 &                    0.769354 &          1.402426 &          -0.275388 &          -0.269869 & 15.174 & $\checkmark$ &      SG \\
J175243.07-285138.2 & 2231.401770 &  3.796457 &           0.290694 &                           0.926136 &  0.046 &  271.465258 & 80.938546 &           &                             &          0.920920 &          -0.013677 &           0.389648 & 11.930 &              &         \\
J180311.94+321113.9 & 2024.201958 &  3.708114 &           0.187988 &                           0.791176 &  0.043 &   89.791941 & 84.817334 &  3.277965 &                    0.653556 &          0.824116 &           0.607653 &          -0.046921 & 11.965 &              &      MS \\
J093238.91-302546.4 & 2063.625999 &  3.394651 &           0.428208 &                           0.386619 &  0.002 &  114.581324 & 83.110132 &  0.744503 &                    0.242247 &          0.758643 &           0.248488 &           0.167120 & 12.892 &              &      MS \\
J071101.65+301928.8 & 2021.248271 &  2.774759 &           0.300369 &                           0.993818 &  0.003 &  118.644627 & 87.296866 &  1.818594 &                    0.468822 &          1.383011 &           1.621214 &           0.395588 & 12.318 &              &      MS \\
J090616.35-382316.0 & 2283.503354 &  2.020317 &           0.423819 &                           0.774946 &  0.000 &  280.124618 & 79.363199 &  2.118347 &                    0.634641 &          0.553822 &           0.047535 &           0.946984 & 12.876 &              &      MS \\
J083938.77-050614.1 &  622.357295 &  1.758010 &           0.382207 &                           0.964809 &  0.058 &   89.716158 & 89.126944 &  3.395996 &                    0.898719 &          1.311298 &           0.331991 &          -0.135843 & 15.327 & $\checkmark$ &      SG \\
J065612.80-291300.2 & 1422.417468 &  1.583792 &           0.594658 &                           0.587456 &  0.000 &   98.239004 & 72.886481 &  1.515521 &                    0.513772 &          2.533157 &          -0.328507 &           0.304922 & 13.908 & $\checkmark$ &      MS \\
J232812.76-395523.9 & 2015.020261 &  0.768702 &           0.437851 &                           0.634919 &  0.000 &  358.903857 & 80.666328 &  2.644607 &                    0.281897 &          0.872177 &          -0.444428 &           0.231803 & 13.185 &              &      MS \\
\bottomrule
\end{tabular}

    \end{center}
\end{table*}

Although this approach is generally effective at producing reliable fits with minimal manual intervention, we visually inspect all \nEA{} light curve solutions to identify systems that require additional optimization or have found a local minimum. In addition to evaluating the $\chi^2$ of the solution, we consider the alarm statistic defined by \citet{Tamuz06}. The phased model residuals are grouped into $M$ `runs' over which consecutive residuals in phase have the same sign and
\begin{equation} \label{eqn:alarm}
    \mathcal{A} = \frac{1}{\chi^2}\sum_{i=1}^{M}\left(\frac{r_{i,1}}{\sigma_{i,1}}+\frac{r_{i,2}}{\sigma_{i,2}}+...+\frac{r_{i,k_i}}{\sigma_{i,k_i}} \right)^2 - \left(1+\frac{4}{\pi}\right)
\end{equation}
where $r_{i,j}$ is the $j$th residual in the $i$th run with uncertainty $\sigma_{i,j}$. The alarm statistic is particularly useful for identifying cases where the eclipse depth or width are poorly fit. We find some cases where the optimization falls into local minima with high eccentricities or low inclinations, in which case we modify the initial conditions and reoptimize the model. While the BLS estimates from Section \S\ref{sec:bls} are generally successful in refining the orbital period from the $V$-band catalog estimates, we find some light curves where the $V$- and $g$-band light curves have small ($\lesssim 0.05$) phase offsets. For these systems we run an additional optimization including $P$ and \tsupconj{} as free parameters. The sigma clipping treatment described in Section \ref{sec:models} almost never over-clips the eclipses, but there were a small number ($\sim$100) of systems that needed to be refit with the sigma-clipping removed.

During visual inspection we remove \nPOOR{} light curves from our catalog. Figure \ref{fig:class_probability} shows that many of these have low classification probabilities from \citet{JayasingheII} with \LowCPPoor{} systems having a classification probability $P_{\rm{class}}<0.9$. We also remove EBs that have poor ASAS-SN light curves, either because they are faint ($g\gtrsim15$~mag) or have poor sampling given the orbital period. These systems are many of the high $\Delta P$ population shown in Figure \ref{fig:bls}. Figure \ref{fig:panel_poor_ea} shows five examples of $g$-band light curves that were removed during visual inspection. 
As part of this process we simultaneously inspect the \textit{TESS} light curves from the SPOC \citep[Sectors 1--38,][]{Caldwell20} and the QLP \citep[Sectors 1--29,][]{Huang20a, Huang20b, Kunimoto21} pipelines. Out of the \nEA{} in our initial catalog, \nQLP{} have QLP light curves for at least one sector and \nSPOC{} have SPOC light curves for at least one sector. Both \textit{TESS} pipelines produce both "raw" and detrended light curves. We examined the "raw" light curves because the detrending can remove real stellar variability. The high cadence \textit{TESS} light curves help to identify shallow eclipses that can be missed in the ASAS-SN light curves. If necessary, we update the period and reoptimize the model. Figure \ref{fig:ea_half_p} shows an example (ASASSN-V J010018.84+552507.1) where the the shallow eclipse is missed in the ASAS-SN data leading to an initial period that is really $P/2$.

We also identify systems with additional physics, such as spots, accretion, and potential triple systems that require more complex models. These \nFLAG{} targets will be the subject of a subsequent paper. We also crossmatched these targets with the ATLAS all-sky stellar reference catalog \citep[ATLAS REFCAT2, ][]{Tonry18} to see if a large fraction of these are blends from the large TESS apertures. The proximity statistic {\tt r1} gives the radius (in arcseconds) where the cumulative flux from nearby stars equals the flux of the target and is set to 99.9 if this value is not reached within 36\arcsec. We find that \nFlagBlends{} of the \nFLAG{} systems have ${\tt r1} < 99.9$. This is consistent with the rest of the catalog, where \percentCleanBlends{}\% of targets have a star within 36\arcsec, suggesting that only a small fraction of the more complex \textit{TESS} light curves are be blends. Appendix \ref{appendix:LCs} includes example light curves for targets flagged as requiring additional physics. 

Many of the systems identified as having long term trends due to rotational variability in Section \S\ref{sec:rot} also show corroborating evidence of rotational variability in the \textit{TESS} light curves. In most cases, the amplitude of the rotational variability is small enough that the \PHOEBE{} model is able to accurately determine the system parameters using the $V$-band light curve. For \ndropg{} systems, we fit only the $V$-band light curve, as the long-term trends in the $g$-band light curve prevent any meaningful contribution to the model solution. 
 
\subsection{Evolutionary States} \label{sec:cmd_methods}

Out of the \nSOL{} in our catalog, \nGaiaTargetsSol{} are in Gaia Early Data Release 3 \citep[EDR3, ][]{GaiaCollaboration16, GaiaCollaboration21}. We apply a quality cut and select systems with {\tt parallax\_over\_error~$>10$}, parallaxes $\pi > 0$, $G$, $G_{\rm{BP}}$, and $G_{\rm{RP}}$ measurements, and distance estimates from \citet{BailerJones21}. We use extinction estimates from the {\tt mwdust} \citep{Bovy16} 3-dimensional `Combined19' dust map \citep{Drimmel03, Marshall06, Green19} and only keep systems with $A_V < \avcutoff{}$~mag, which roughly corresponds to the 95th percentile of the distribution. We use Table 3 of \citet{Wang16} to convert to $A_G$ and $E(G_{\rm{BP}}-G_{\rm{RP}})$. In total, \nGaiaTargetsFilteredSol{} EBs have extinction-corrected absolute magnitudes and colors. 

Since the eclipse probability increases with radius, we expect a relatively high fraction of subgiants despite the shorter timescale of the evolutionary state. Given the CMD, we separate the main sequence, subgiant, and red giant stars using the MESA Isochrones \& Stellar Tracks \citep[MIST,][]{Choi16, Dotter16} isochrones. We define the subgiant branch to begin with at the terminal age main sequence (TAMS). We interpolate over Solar metallicity MIST isochrones ranging in age from $10^8$ to $10^{10}$ years in intervals of $0.1$ in dex to define this boundary. To represent binary star isochrones, we double the flux in each band to represent a binary of equal mass. We define the subgiant branch to end when the radius $R=1.5 R_{\rm{TAMS}}$, where $R_{\rm{TAMS}}$ is the radius at the terminal age main sequence. We interpolate over MIST evolutionary tracks for masses 1--6~M$_\odot$ to define this boundary. We set the maximum absolute magnitude limit for the subgiant/giant branch to be at $M_G=4.5$~mag.

\section{Distribution of Eclipsing Binary Parameters} \label{sec:results}

\begin{figure*}
    \centering
    \includegraphics[width=\linewidth]{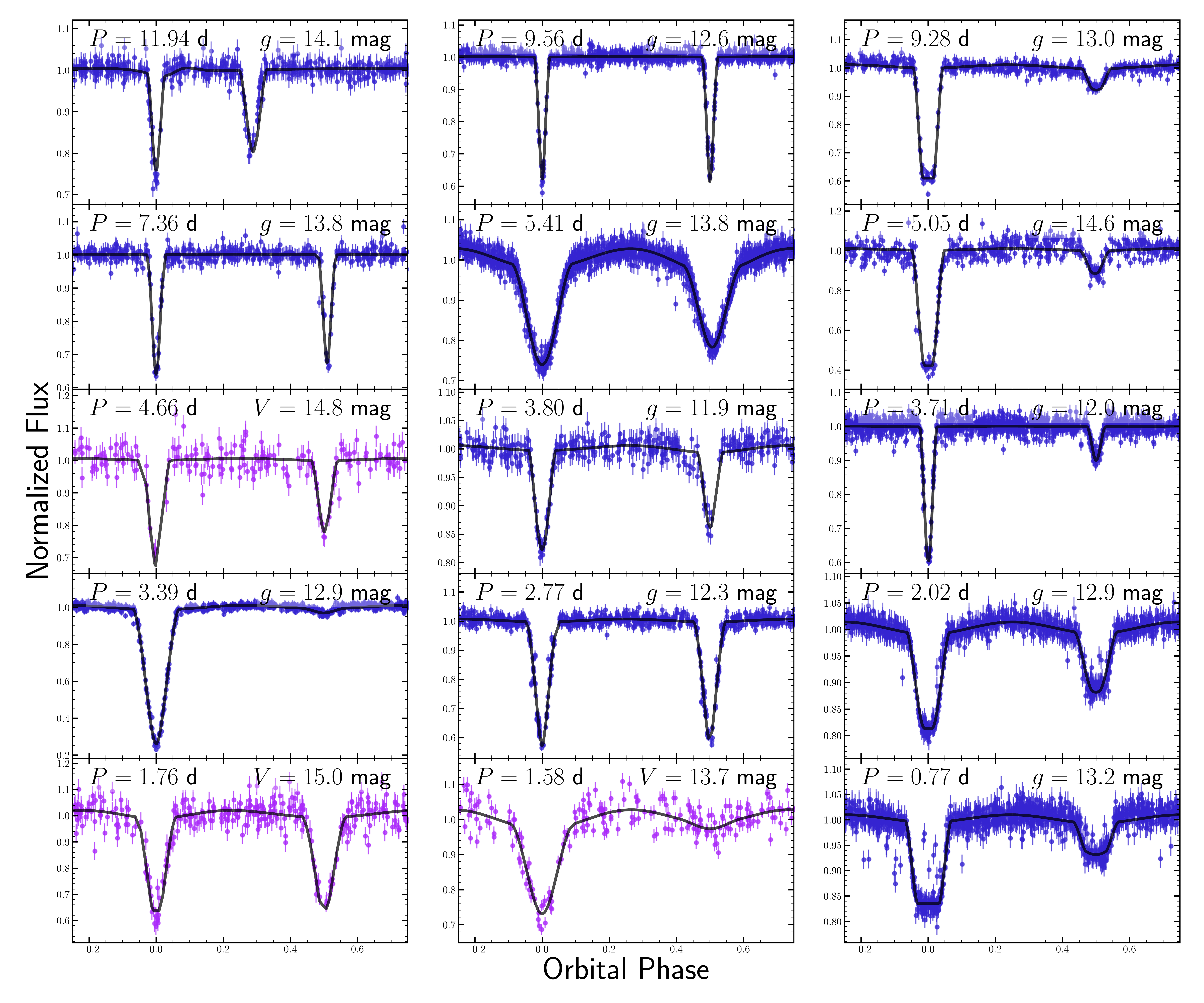}
    \caption{ASAS-SN $V$- (purple) and $g$-band (blue) light curves for the eclipsing binaries in Table \ref{tab:ea_table}. Systems with long-term trends from systematic effects for bright sources or rotational variability are shown with their $V$-band light curves. The optimized models are shown in black. The phase range of $-0.25$ to $+0.75$ is used to clearly show both eclipses.}
    \label{fig:panel_plot}
\end{figure*}

\begin{figure*}
    \centering
    \includegraphics[width=\linewidth]{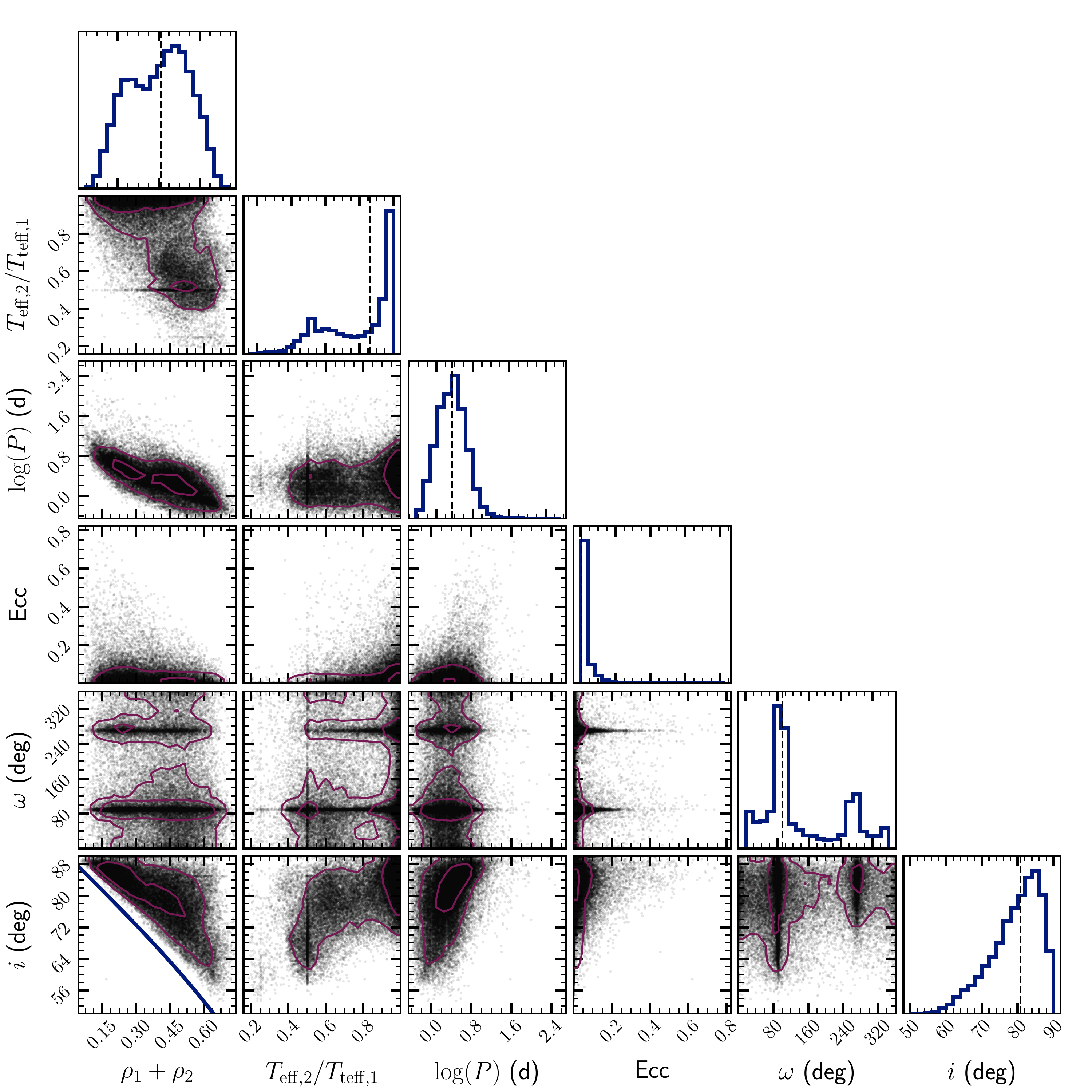}
    \caption{The distributions of the fitted parameters. The vertical lines show the median for each distribution. Contours are shown for each distribution, and the bottom left panel shows the inclination limit for detecting systems as a function of \requivsumfrac{}.}
    \label{fig:corner}
\end{figure*}

The distribution of the EB model parameters from the \PHOEBE{} models can be used to explore both the intrinsic binary parameter distributions as well as selection effects in our sample. Table \ref{tab:ea_table} gives model parameters for the \nSOL{} eclipsing binaries remaining after visual inspection. A selection of the light curve solutions are shown in Figure \ref{fig:panel_plot} and in Appendix \ref{appendix:LCs} and Figure \ref{fig:corner} shows their parameter distributions. The features of these distributions are combinations of probabilistic, systematic, and physical features.

\subsection{Inclination and Argument of Periastron Distribution} \label{sec:i_w}
The distribution of inclinations is peaked toward edge-on systems, and the lowest inclination systems are almost exclusively found at short periods and high \requivsumfrac{}. This is simply due to the eclipse probability scaling as $\sim(\rho_1+\rho_2)P^{-2/3}$. Eccentric systems also tend to be more edge on because it is easier to miss one of the eclipses for inclined systems.

The distribution of the argument of periastron $\omega$ has two peaks at $90^{\circ}$ and $270^{\circ}$. For circular orbits where the phase separation of the eclipses is 0.5, the geometry estimator sets the argument of periastron to $\omega=90^{\circ}$. While this parameter sometimes deviates from this initial value during optimization, usually flipping by $180^{\circ}$, it is poorly constrained unless the binary is significantly eccentric. 

\subsection{Eccentricity Distribution} \label{sec:period_ecc}

\begin{figure*}
    \centering
    \includegraphics[width=\linewidth]{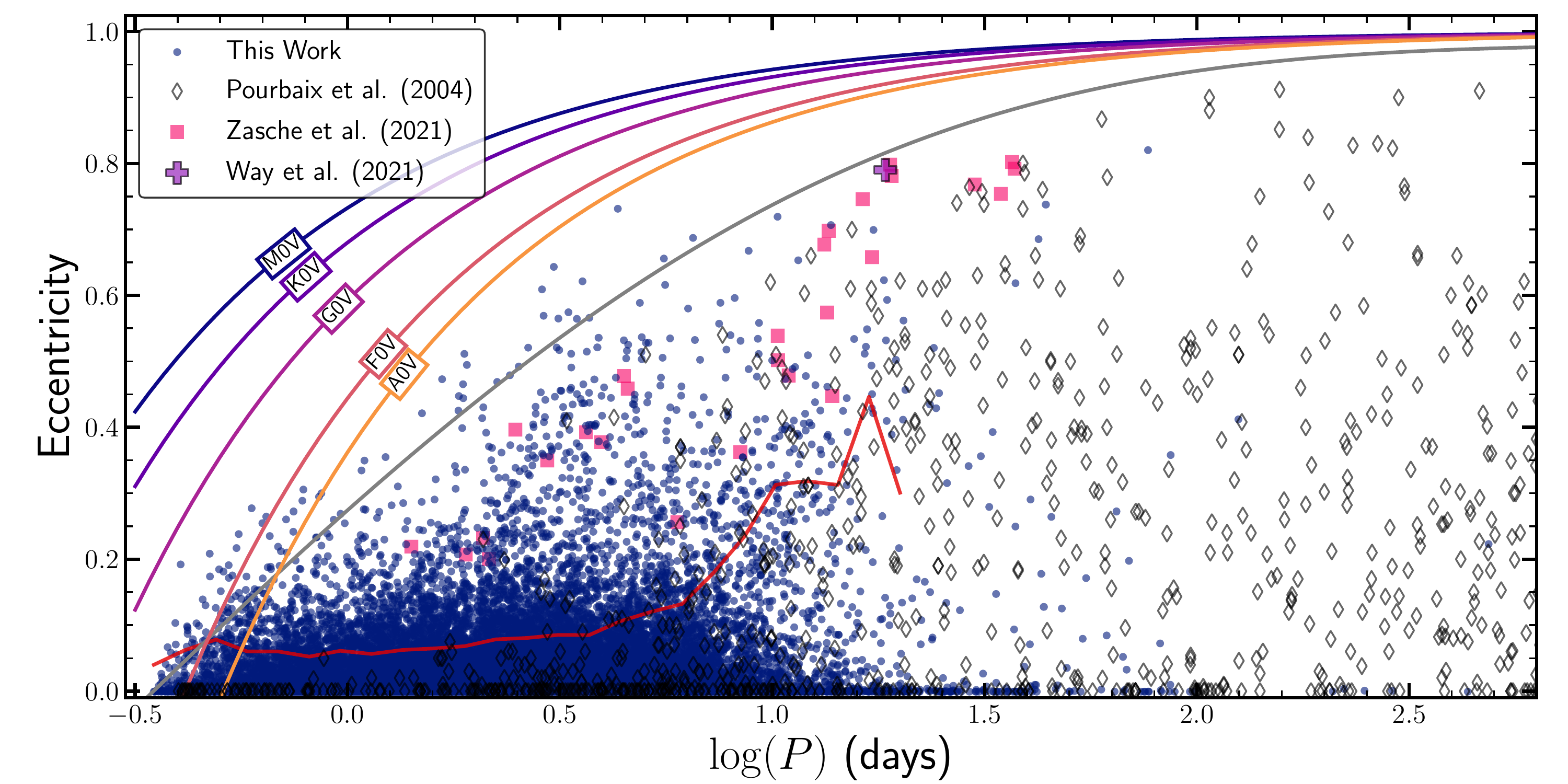}
    \caption{The distribution of periods and eccentricities. We compare the ASAS-SN sample (blue dots) to a catalog of eccentric EBs identified by \citet[pink squares,][]{Zasche21} and to the spectroscopic binaries from the SB9 catalog \citep[black diamonds,][]{Pourbaix04} with $\rm{Grade}\geq4$. The gray line shows the eccentricity envelope found by \citet{Mazeh08} for the SB9 catalog. The colored lines show the eccentricities where binaries of equal mass would be in contact at pericenter for different spectral types. The red line shows the 90th quantile of the eccentricity of bins in $\log P$.}
    \label{fig:period_ecc}
\end{figure*}

Figure \ref{fig:period_ecc} shows the distribution of EB periods and eccentricities along with the SB9 catalog of spectroscopic orbits \citep{Pourbaix04} and the \citet{Zasche21} catalog of eccentric eclipsing binaries for comparison. The extremely high eccentricity ($e=0.79$) system identified in \citet{Way21} is also labeled. We find that \PercentCirc{}\% of systems have eccentricities consistent with circular orbits ($e<0.05$). The spread in eccentricity expands with increasing periods until $\sim10$~d, after which few $e>0$ systems are observed. The increasing absence of short period eccentric systems is due to tidal circularization \citep{Verbunt95}

Figure \ref{fig:period_ecc} shows the upper eccentricity envelope of
\begin{equation} \label{eqn:MazehEnvelope}
    e < E-A\exp{(-(PB)^c)}
\end{equation}
with $E=0.98$, $A=3.25$, $B=6.3$, and $C=0.23$ found by \citet{Mazeh08} for the SB9 catalog. Our parameters are generally consistent with this envelope, but some lie beyond this empirical model. We also show (very) hard upper limits for the eccentricities of binary twins of different spectral types using the eccentricity $e=(1-2 R_*/a)$ where the stars would be in contact at pericenter where $a$ is the semi-major axis and values of $M_*$ and $R_*$ are taken from \citet{Pecaut13}. Systems that fall above the \citet{Mazeh08} envelope are still consistent with the range of acceptable eccentricities for main sequence binaries of spectral type A0V.

In general, we expect to find fewer high eccentricity systems in eclipsing systems than in spectroscopic binaries because at a fixed period, the probability of detecting a second eclipse decreases with increasing eccentricity. When only the first eclipse is detected the system will tend to be modeled either as a system with \teffratio{}$\sim1$ at half of the true period or with \teffratio{}$\ll1$ at the correct period. In total, we find \nEccpI{} systems with $e>0.1$, \nEccpIIV{} systems with $e>0.25$, and \nEccpV{} systems with $e>0.5$. Figure \ref{fig:panel_high_ecc} shows examples of light curves with high eccentricity and Figure \ref{fig:panel_short_p_high_ecc} shows examples of short period systems with non-zero eccentricities. 

\subsection{Period Distribution} \label{sec:period_distribution}

\begin{figure*}
    \centering
    \includegraphics[width=0.8\linewidth]{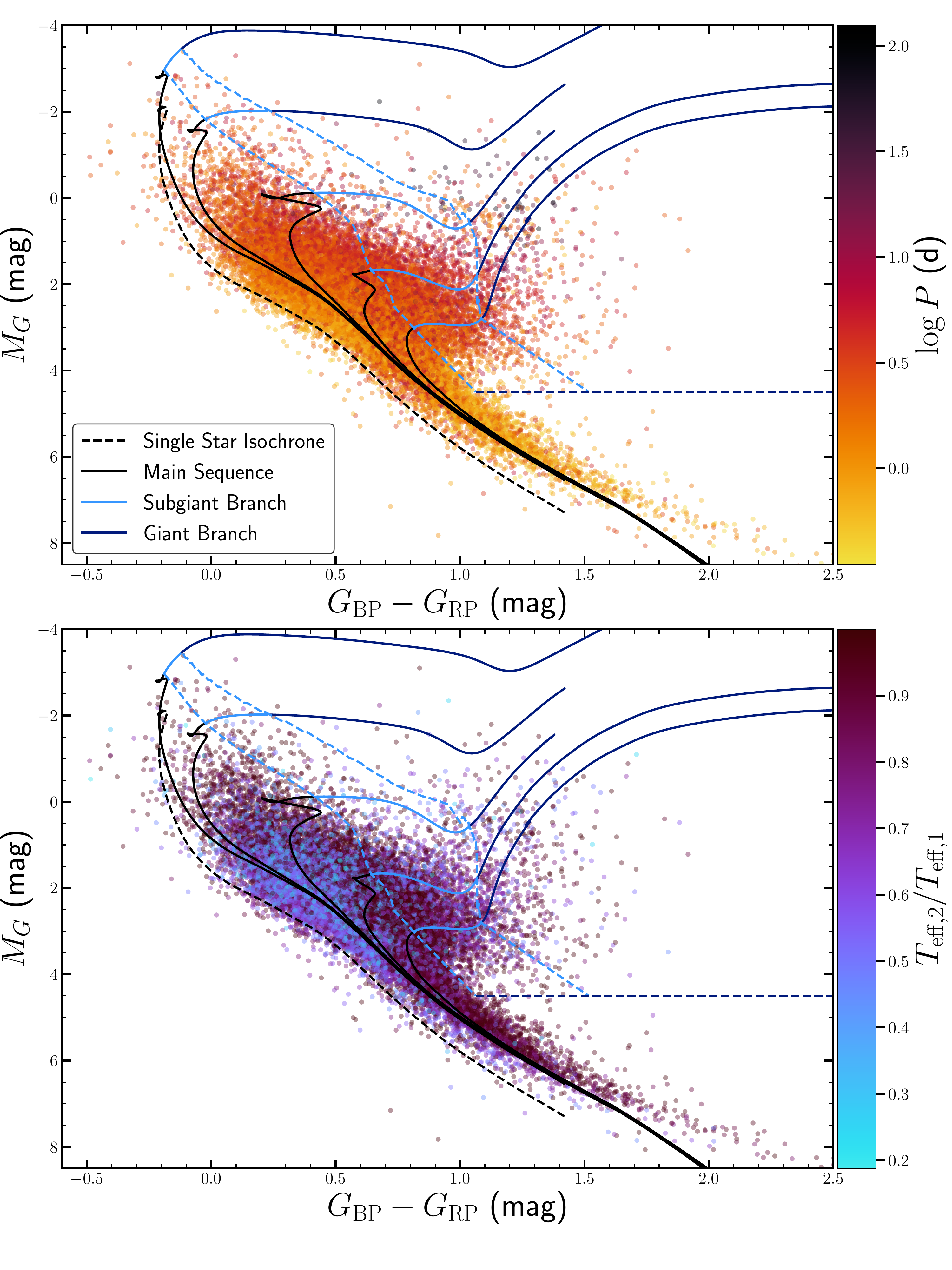}
    \caption{Detached EBs on a Gaia EDR3 color-magnitude diagram (CMD) after removing systems with {\tt parallax\_over\_error} $<10$ or $A_V<\avcutoff{}$~mag. The EBs are colored by $\log P$ (top) and \teffratio{} (bottom). The solid lines show the MIST isochrones that are used to determine the divisions of evolutionary states (blue dashed lines) for binaries of equal mass. For comparison, a single star isochrone is shown as the black dashed line.}
    \label{fig:CMDs}
\end{figure*}

\begin{figure*}
    \centering
    \includegraphics[width=\linewidth]{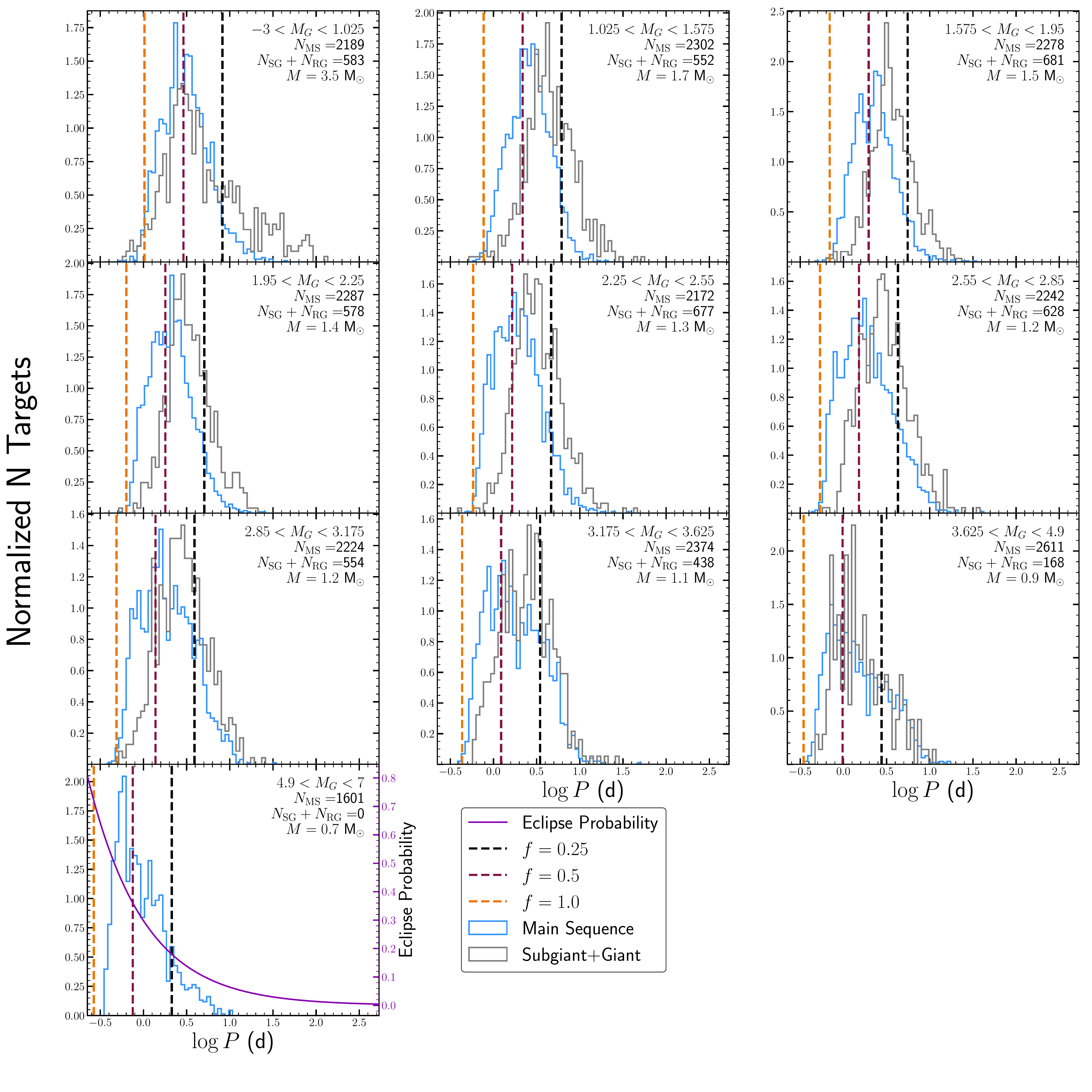}
    \caption{Distribution of orbital periods for different absolute magnitude bins ordered from brightest to faintest magnitude bin from left to right then top to bottom. The solid blue histograms show the main sequence primaries selected using the MIST isochrones and the gray histograms show the subgiant/red giant systems. The absolute magnitude $M_G$ range, number of main sequence primaries, $N_{\rm{MS}}$, number of subgiant+red giant primaries, $N_{\rm{SG}}+N_{\rm{RG}}$, and main sequence mass $M$ are included for each magnitude bin. The dashed vertical lines show the periods where equal mass binaries have a ratio $f=R/R_{\rm{RL}}$ between the stellar and Roche radii. Finally, the purple line in the faintest magnitude bin shows the $P^{-2/3}$ eclipse probability for main sequence stars (Equation \ref{eqn:eclipse_prob}).}
    \label{fig:CMD_binned_periods}
\end{figure*}

Out of the \nSOL{} detached eclipsing binaries in our catalog after visual inspection, \nGaiaTargetsSol{} are in Gaia EDR3 and \nGaiaTargetsFilteredSol{} meet the quality cuts described in Section \ref{sec:cmd_methods}. Figure \ref{fig:CMDs} shows the Gaia EDR3 extinction-corrected color magnitude diagram with systems colored by $\log(P)$. A large fraction of long period systems are found for higher mass main sequence stars and evolved stars partly because shorter period systems increasingly cannot exist around these stars and because the detection probability for long period systems is smaller for lower mass, smaller radius stars. 

The CMD suggests that there is a large population of subgiant primaries in the EB catalog. Using the MIST isochrone divisions of the CMD as described in Section \ref{sec:cmd_methods}, we identify \nPrimaryMS{} main sequence primaries, \nPrimarySG{} subgiant primaries, and \nPrimaryRG{} giant primaries (Table \ref{tab:ea_table}). The full orbital period distribution contains contributions from short period systems with low mass primaries, early-type stars in longer period binaries, and giants in the longest-period systems. To better understand the period distribution, we divided the CMD into 13 absolute magnitude bins each containing $\sim 2750$ stars with ${\tt parallax\_over\_error}>10$ and $A_V < \avcutoff{}$~mag, and use MIST isochrones to separate the main sequence stars from the subgiants and giants as described in Section \S\ref{sec:cmd_methods}. Figure \ref{fig:CMD_binned_periods} shows the orbital period distribution for each magnitude bin. 

The bottom panel of Figure \ref{fig:CMD_binned_periods} shows the eclipse probability for twin main sequence binaries of mass $M$ and radius $R$, 
\begin{equation} \label{eqn:eclipse_prob}
    \mathcal{P} = 2 R \left(\frac{2GM}{4\pi^2}\right)^{-1/3} P^{-2/3}.
\end{equation}
To determine $M$ and $R$ for each $M_G$ bin, we use a grid of Solar-metallicity evolutionary tracks from MIST and integrate over the main sequence lifetime to get the average $M_G$ and $R$. We then interpolate the relations between $M_G$--$M$ and $M_G$--$R$ to solve for the typical stellar parameters of the $M_G$ bin. As before, we double the Gaia $G$-band flux to consider equal mass binaries. 

The vertical lines in Figure \ref{fig:CMD_binned_periods} show the period, \begin{equation}
    P_{\rm{Roche}}(f) = \sqrt{\frac{4\pi^2}{2 G M}\left(\frac{R/f}{E(q)}\right)^3}
\end{equation}
corresponding to the Roche radius where $f$ is the Roche-lobe filling factor and
\begin{equation}
    E(q) \equiv \frac{0.49 q^{-2/3}}{0.6 q^{-2/3} + \ln(1+q^{-1/3})}.
\end{equation}
is the \citet{Eggleton83} approximation for the scaling with mass ratio $q$. We compute these periods for binaries of equal mass, taking $M$ and $R$ from MIST evolutionary tracks, as before. When moving to fainter magnitude bins, the distribution of orbital periods moves to shorter periods, reflecting the lower detection probability for low-mass stars in long-period orbits and the minimum period required for a detached system at the high mass end. As expected, systems with subgiant/giant primaries are found at larger orbital periods than main sequence primaries of similar absolute magnitude $M_G$.

\subsection{Effective Temperatures and Fractional Radii} \label{sec:teffratio_and_radii}

\begin{figure*}
    \centering
    \includegraphics[width=\linewidth]{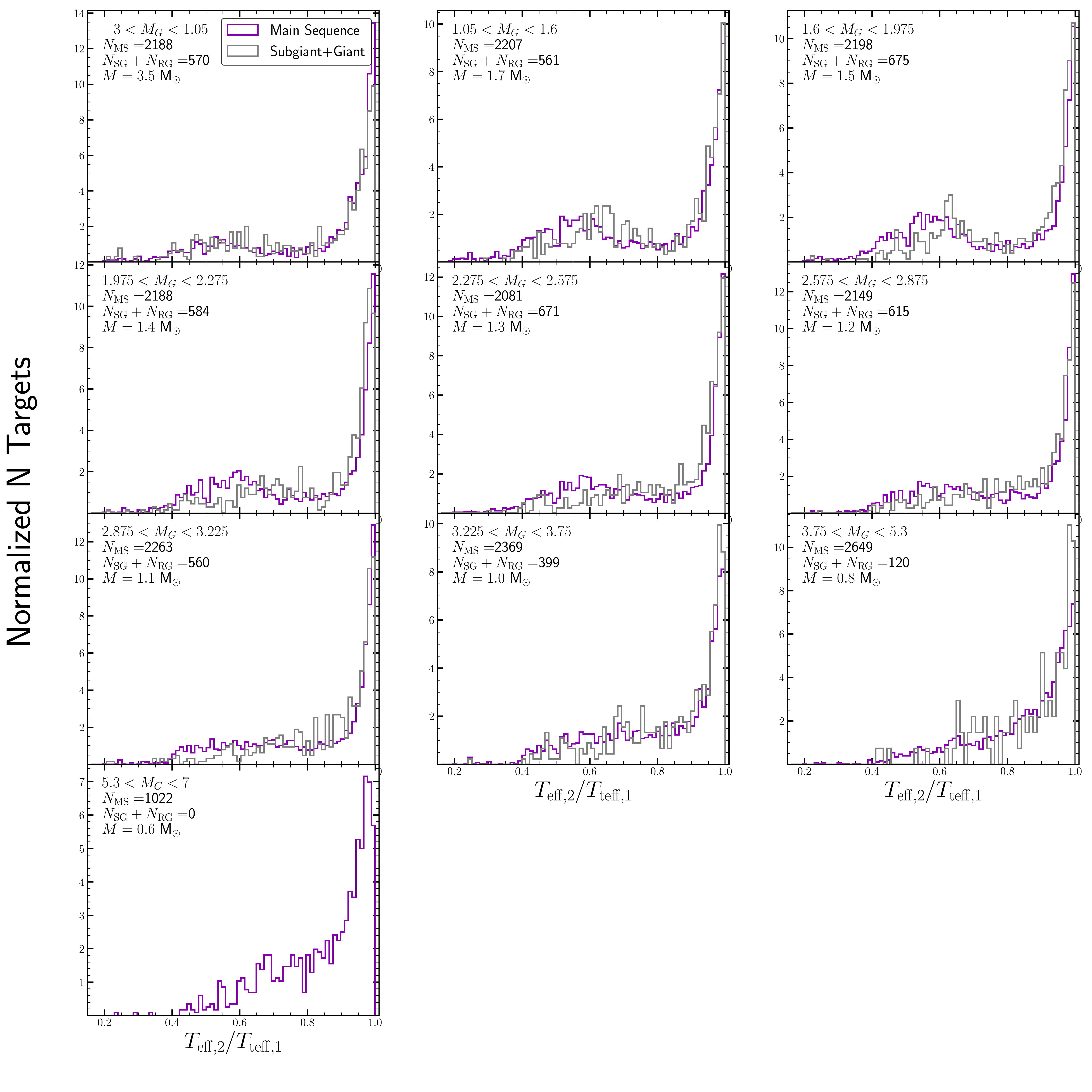}
    \caption{Same as Figure \ref{fig:CMD_binned_periods}, but for the \teffratio{}. The total distribution of \teffratio{} shown in Figure \ref{fig:corner} shows three components: a modeling artifact at \teffratio$\sim0.5$, a sharp peak at \teffratio{}$\sim1.0$, and a broad Gaussian near \teffratio{}$\sim0.6$. The relative amplitude of this broad component decreases with increasing $M_G$, reflecting the tendency of low \teffratio{} systems to be found higher on the CMD (Figure \ref{fig:CMDs}).}
    \label{fig:CMD_binned_teffs}
\end{figure*}

\begin{figure*}
    \centering
    \includegraphics[width=0.75\linewidth]{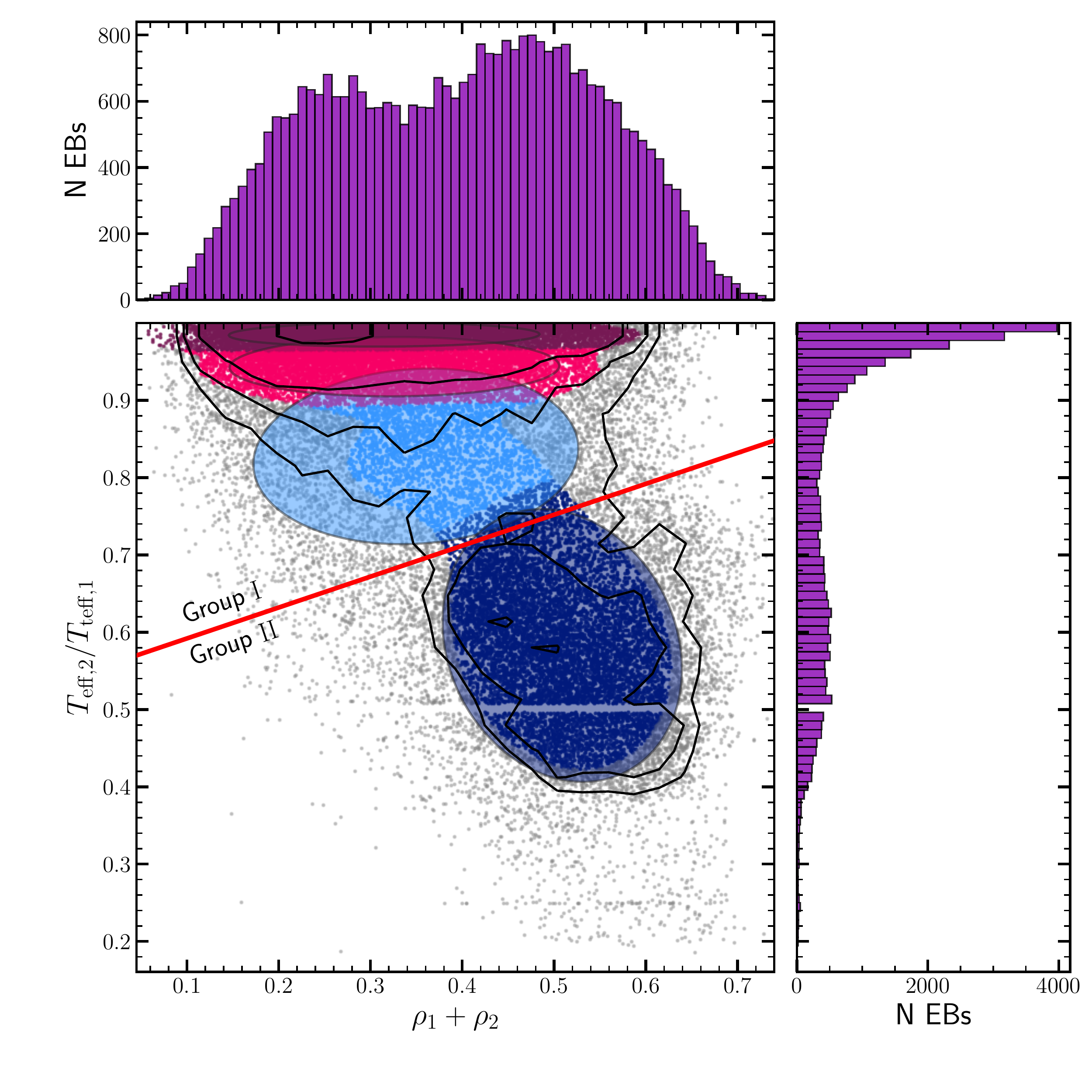}
    \caption{Distribution of \requivsumfrac{} and \teffratio{} with systems in the modeling artifact near \teffratio{}$\sim0.5$ removed. The colored ellipses show the components of the Bayesian Gaussian mixture model. Colored points show the systems with the highest log-likelihood of being associated with each component. We use this model to roughly divide the systems into Group I (high \teffratio{}) and Group II (low \teffratio{}) using the red line defined in Equation \ref{eqn:groupIgroupII_line}.}
    \label{fig:groupIgroupII_scatter}
\end{figure*}

\begin{figure}
    \centering
    \includegraphics[width=\linewidth]{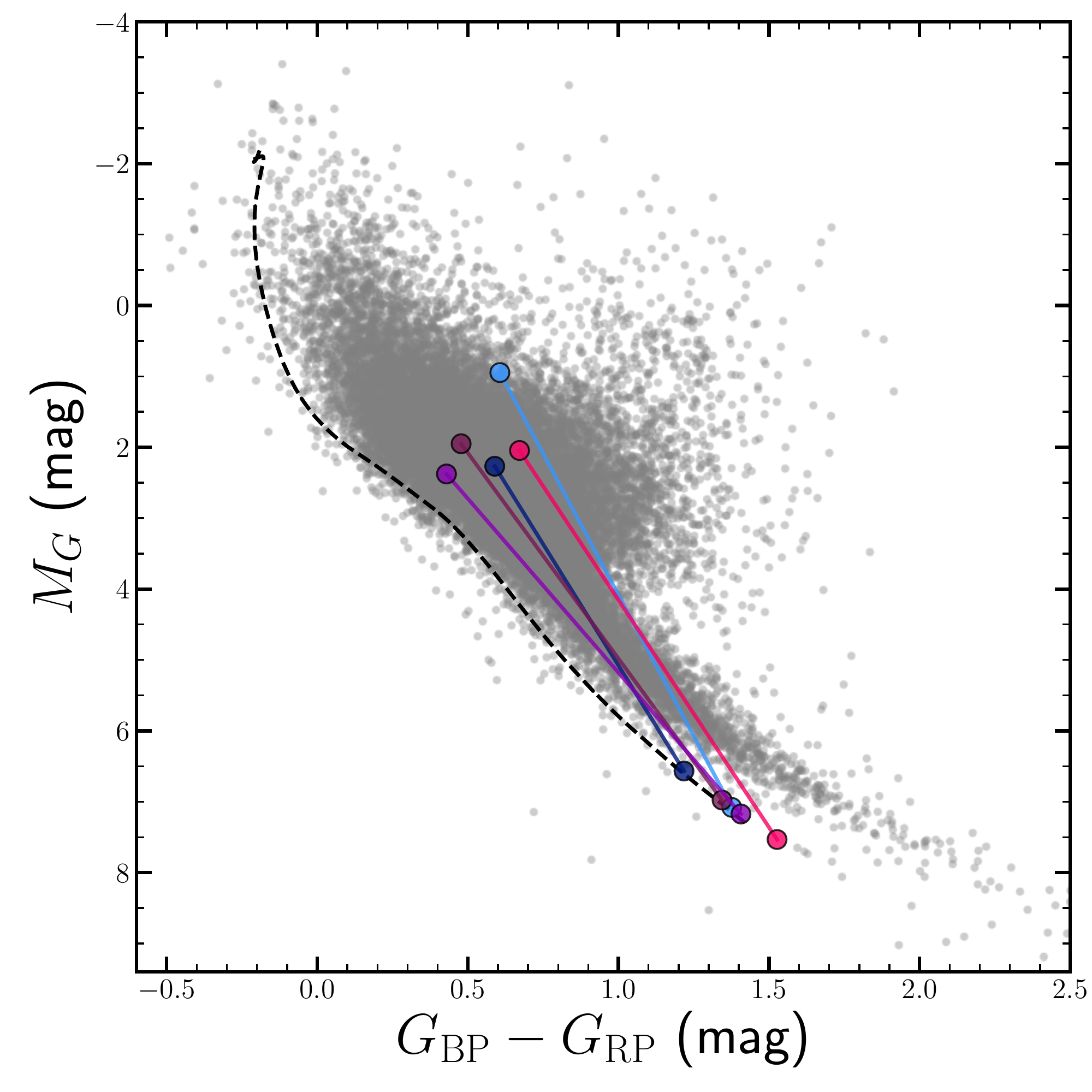}
    \caption{We combine the extinction-corrected Gaia magnitudes, the fit light curve parameters, and MIST evolutionary tracks to estimate the properties of the components of example Group II binary systems. Each binary system is represented by a pair of points, one for each component, connected by a line of the same color. The black line shows a single-star isochrone. The light curves for these systems are shown in Figure \ref{fig:panel_groupII}.}
    \label{fig:groupIgroupII_separate_cmd}
\end{figure}

\begin{figure}
    \centering
    \includegraphics[width=\linewidth]{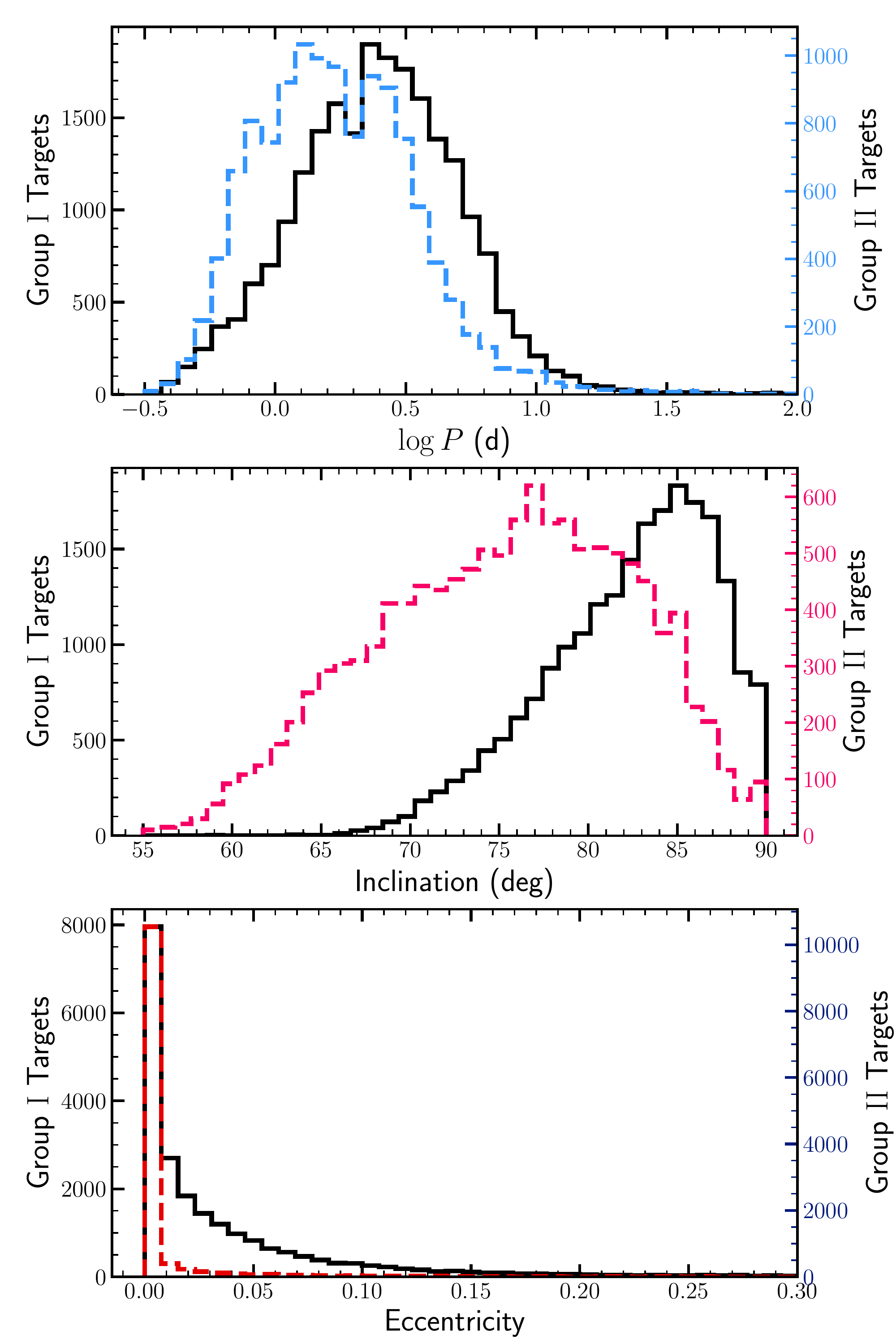}
    \caption{Distributions of $\log P$, inclination, and eccentricity for EBs in Group I (black histograms), Group II (colored histograms). We find that Group II EBs are typically at shorter periods, lower inclinations, and are in more circular orbits. }
    \label{fig:groupIgroupII_hists}
\end{figure}

\begin{figure}
    \centering
    \includegraphics[width=\linewidth]{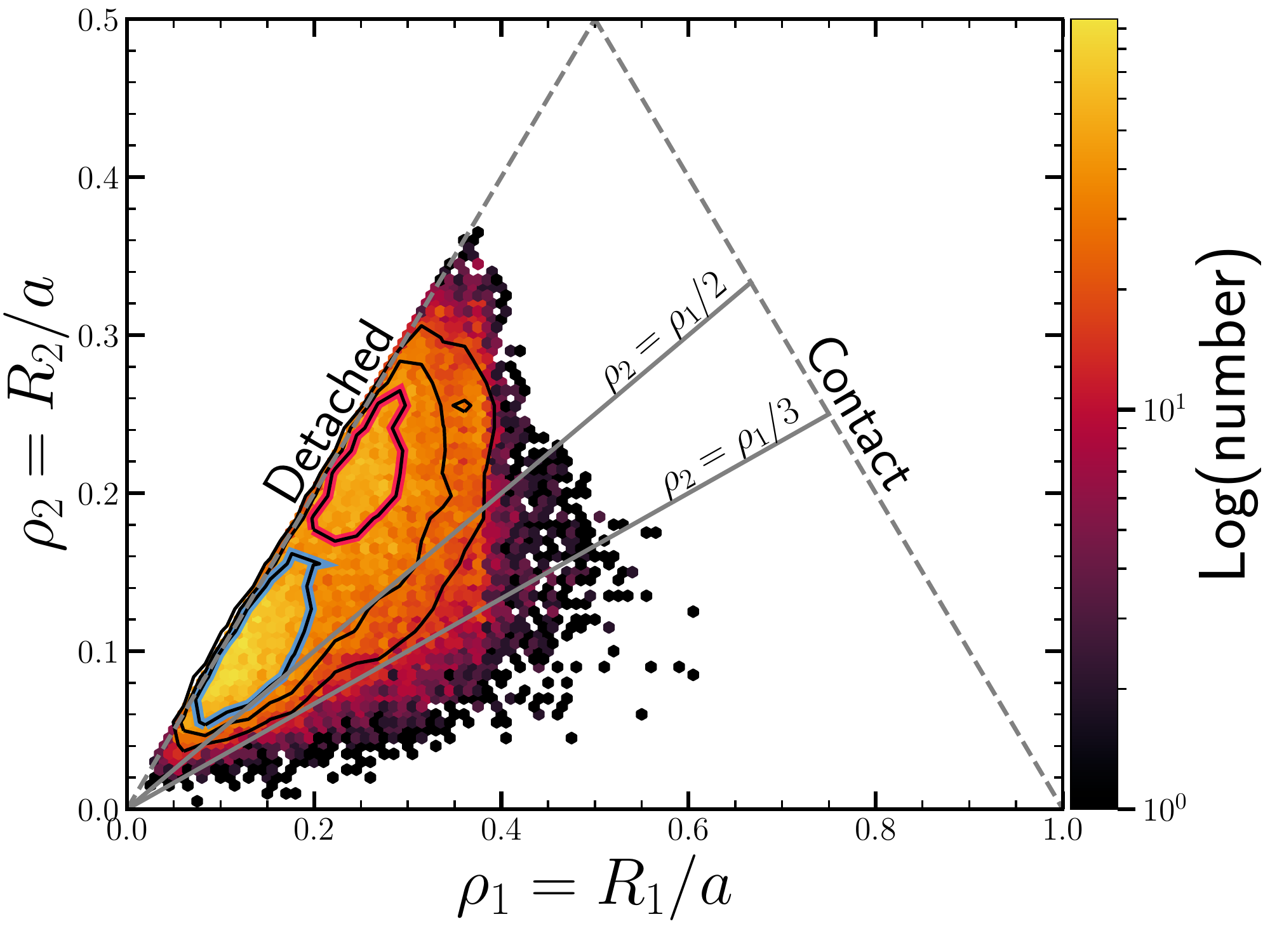}
    \caption{Fractional radii for each component of the eclipsing binary where $\rho_1$ is defined to be the larger component. The dashed lines indicate the detached limit where $\rho_1=\rho_2$ (left) and the contact limit where $\rho_1+\rho_2=1$ (right). The solid black lines show contours of the distribution, and we find two populations near the detached limit, centered $\rho_1\sim0.15$ and $\rho_1\sim0.25$ (highlighted contours).}
    \label{fig:radius_radius}
\end{figure}

There are three features in the \teffratio{} and \requivsumfrac{} distributions. The linear feature at \teffratio{}$=0.5$ is a modeling artifact from systems with no detectable secondary eclipse. The rest of the systems form two distinct groups. One group is concentrated at roughly equal temperatures and smaller fractional radii, while the second group has larger temperature differences and larger fractional radii. The second group has shorter periods, which is not surprising given the larger fractional radii. Figure \ref{fig:groupIgroupII_scatter} shows an expanded view of this distribution with the systems that fall in the linear feature at \teffratio$=0.5$ removed.
\begin{equation} \label{eqn:groupIgroupII_line}
    T_{\rm{eff},2}/T_{\rm{eff},1} = 0.4(\rho_1+\rho_2)+0.55
\end{equation}
roughly separates the two populations and we label systems that fall above and below the line as Groups I and II, respectively. The light curves in Group I are ``classical'' detached binaries, while the Group II light curves resemble semidetached systems with more curved out-of-eclipse shapes. Examples of Group II light curves are shown in Figure \ref{fig:panel_groupII}. 

Figure \ref{fig:CMDs} shows the EBs on a Gaia CMD colored by \teffratio{}. Systems with lower \teffratio{} are preferentially found on the upper main sequence and below the isochrones corresponding to the equal mass binaries. This is seen more clearly in Figure \ref{fig:CMD_binned_teffs} where we show the distribution of \teffratio{} in different $M_G$ bins after removing systems with \teffratio{}$\sim$0.5. The broad Gaussian feature corresponding to the low \teffratio{} Group II systems decreases moving down the CMD, more rapidly for the evolved stars than the main sequence stars, and disappears entirely by $M_G\sim 3.5$~mag. Lower main sequence stars span a broad range of luminosities but a limited range of temperatures, so lower main sequence stars with the temperature ratios of Group II will have large differences in luminosity and size, making it difficult to detect eclipses. And, at the very bottom of the MS, it is simply not possible to have a companion with half the temperature of the primary. 

We can investigate the likely properties of the stars by taking pairs of stars from MIST isochrones with primary masses of $M_1=0.5$M$_\odot$ to $8.0$M$_\odot$ and secondary masses $M_2=0.5$M$_\odot$ to $M_1$ and keeping all pairings that match the extinction-corrected Gaia $M_G$ magnitude and $G_{\rm{BP}}-G_{\rm{RP}}$ color of the binary to $0.1$ and $0.01$~mag, respectively. From each of these pairings we find the one that minimizes
\begin{equation}
    \chi^2 = (T_{\rm{eff},2}/T_{\rm{eff},1}-(T_{\rm{eff},2}/T_{\rm{eff},1})_{\rm{model}})^2+(\rho_1+\rho_2-(\rho_1+\rho_2)_{\rm{model}})^2.
\end{equation}
Figure \ref{fig:groupIgroupII_separate_cmd} shows the results for the 5 Group II systems shown in Figure \ref{fig:panel_groupII}. Each seems to pair an upper main sequence star with a low-mass dwarf. 

Finally, Figure \ref{fig:groupIgroupII_hists} shows the distribution of periods, inclinations, and eccentricities for Group I and II systems. Very few of the Group II systems are eccentric, so they must be sufficiently compact to be tidally circularized. More curiously, we find a dearth of edge-on Group II systems. With finite errors on the inclination estimate, we expect the distribution to drop as the inclination approaches edge-on, as seen for the Group I systems. However, the Group II distribution is already falling by $\sim 75^\circ$. The lack of high-inclination Group II systems could be due to the competing effects of inclination and $T_{\rm{eff},1}$ in determining the eclipse depth. In standard detached eclipsing binaries it is the ratio of the temperatures, not the absolute temperatures, that dictates the eclipse depth. However, the ellipsoidal variations observed in Group II light curves are dependent on the temperature-dependent limb-darkening parameters. As a result of fixing the primary effective temperature in Section \ref{sec:models}, the inclination distribution may be artificially skewed to lower inclinations. In addition, it is also possible that the high-inclination semi-detached systems were classified as semidetached rather than detached in \citet{JayasingheII} and thus were not included in our catalog.

Although the individual fractional radii are poorly constrained for most systems except for total eclipses and eccentric orbits, and we only give the sum of the fractional radii for the \PHOEBE{} model solutions in Table \ref{tab:ea_table}, we show the fractional radii for each component in Figure \ref{fig:radius_radius}. As expected, the distribution is skewed to the left boundary because we examined only detached systems. We find that there are two populations along the boundary, near $\rho_1\sim0.15$ and $\rho_1\sim0.25$. A similar distribution was observed by \citet[][Fig. 10]{Devor05}, who suggested that the clustering is due to the systems in the lower $\rho_1$ cluster having their periods erroneously doubled. While it is unlikely that so many systems had inaccurate periods after visual inspection, we inspected $\sim$500 TESS light curves for targets in the smaller $\rho_1$ cluster and confirm that all the periods were correct. We find that the smaller $\rho_1$ cluster consists of Group I systems from the left peak of the \requivsumfrac{} distribution (Figure \ref{fig:groupIgroupII_scatter}). The \nClumpUpper{} systems in the highlighted upper cluster in Figure \ref{fig:radius_radius} are evenly divided between Group I and Group II systems. 

\subsection{High Energy Emission} \label{sec:high_energy_emission}

High energy emission produced from chromospheric or coronal activity is closely related to stellar rotation and magnetic fields \citep{Walter81, Dobson89, Pizzolato03}. Eclipsing binaries with high energy emission can be used to model spot activity in greater detail than for single stars \citep[e.g.,][]{Lanza98} and to study the connection between orbital period modulations and magnetic fields \citep{Applegate92}. Strong stellar winds can also lead to X-ray emission through wind accretion \citep[e.g.,][]{Linder09}. By combining the ASAS catalog of eclipsing binaries \citep{Pojmanski05} with the ROSAT All-Sky Survey \citep{Voges99}, \citet{Szczygiel08} and \citet{Kiraga12} identified 836 and 347 systems, respectively, with evidence of coronal or chromospheric activity. 

We follow a similar procedure and match our catalog of detached eclipsing binaries to the HEASARC Master X-ray catalog\footnote{\url{https://heasarc.gsfc.nasa.gov/W3Browse/all/xray.html}} and the Swift-XRT Point Source Catalog \citep{Evans20}, both with a search radius of $10\farcs0$. We identify \nXRAY{} unique targets with X-ray detections, including \nXRAYMultiple{} targets with multiple detections. Table \ref{tab:xray} presents the EBs with X-ray detections including the X-ray luminosity, $L_X$, computed using the Gaia EDR3 parallax, and the angular separation between the X-ray and optical position. The $N_H$ column density is also given in Table \ref{tab:xray} and is estimated from the {\tt mwdust} extinction and the dust-to-gas ratio $E(B-V)/N_H=(1.5\pm0.5)\times10^{-22}$~mag~cm$^{2}$ from \citet{Dai09}. We did not correct the X-ray luminosities for absorption. In most cases it will be small, and where it is large, the correction depends on the assumed spectrum. For systems included in multiple X-ray catalogs, the flux corresponding to the longest exposure is used to calculate $L_X$. We find that $L_X$ ranges from \lxmin{}~erg/s to \lxmax{}~erg/s with a median value of \lxmed{}~erg/s. The systems with $L_X > 10^{34}$~erg/s are all ROSAT observations with larger offsets that would require additional confirmation. Future and ongoing missions like eROSITA \citep{Predehl21} with positional uncertainties of a few arcseconds could be used to confirm these targets and identify additional eclipsing binaries with X-ray emission.

\begin{table*}
    \centering
    \caption{X-ray detections of detached EBs from Table \ref{tab:ea_table}. The X-ray luminosity is calculated using the observed flux and Gaia EDR3 parallax. The column density $N_H$ is calculated using the {\tt mwdust} extinction and dust-to-gas ratio from \citet{Dai09}. For targets detected by multiple observatories, we include the observation corresponding to the greatest exposure time. The full table is available online.} 
    \label{tab:xray}
    \sisetup{table-auto-round,
             group-digits=false}
    \begin{center}
    \begin{tabular}{lS[table-format=2.2]lrrrS[table-format=2.2]r}
\toprule
     {ASAS-SN Name} &    {$P$} & {Observatory} & {Distance} &              {$L_X$} &              {$N_H$} &  {Separation} &        {Reference} \\
{} & {(d)} & {} & {(pc)} & {(erg/s)} & {(cm$^{-2}$)} & {($\arcsec$)} & {}\\ 
\midrule
J000225.41+701848.3 & 0.523628 &    XMM-NEWTON &      690.0 & $3.0 \times 10^{30}$ & $2.7 \times 10^{21}$ &      2.224406 &     \citet{Webb20} \\
J001856.11-795455.6 & 0.903112 &    XMM-NEWTON &      324.0 & $7.5 \times 10^{30}$ & $3.1 \times 10^{20}$ &      1.455228 &     \citet{Webb20} \\
J002148.55+205512.2 & 1.137296 &         SWIFT &     2309.0 & $4.1 \times 10^{31}$ & $4.6 \times 10^{20}$ &      0.706373 &      \citet{Dai15} \\
J002415.10+603501.1 & 0.549450 &         ROSAT &      147.0 & $7.1 \times 10^{30}$ &                      &      7.157159 &   \citet{Boller16} \\
J002717.28-233642.5 & 1.532329 &         SWIFT &      572.0 & $4.6 \times 10^{30}$ &                      &      2.054949 &    \citet{Evans20} \\
J003328.72+424333.1 & 0.779429 &    XMM-NEWTON &      651.0 & $4.7 \times 10^{30}$ & $2.9 \times 10^{20}$ &      0.264459 & \citet{Traulsen20} \\
J003338.67+392024.0 & 3.330769 &    XMM-NEWTON &      676.0 & $1.5 \times 10^{30}$ & $2.3 \times 10^{20}$ &      1.048870 & \citet{Traulsen20} \\
J003827.07+410330.2 & 6.278181 &    XMM-NEWTON &     1532.0 & $7.7 \times 10^{30}$ & $4.0 \times 10^{20}$ &      1.005259 & \citet{Traulsen20} \\
J003905.81+791911.4 & 1.225147 &         ROSAT &      150.0 & $2.5 \times 10^{31}$ &                      &      6.305507 &   \citet{Boller16} \\
J004138.67+403215.7 & 2.215003 &    XMM-NEWTON &     1203.0 & $3.7 \times 10^{30}$ & $3.5 \times 10^{20}$ &      1.720832 & \citet{Traulsen20} \\
J004409.05-731418.6 & 6.052377 &    XMM-NEWTON &    13135.0 & $4.2 \times 10^{32}$ & $2.5 \times 10^{21}$ &      9.410134 & \citet{Traulsen20} \\
J004459.34-013020.6 & 1.961248 &         SWIFT &     2099.0 & $7.1 \times 10^{31}$ & $2.9 \times 10^{20}$ &      2.261374 &    \citet{Evans20} \\
J004515.37+412523.6 & 1.566936 &    XMM-NEWTON &     2187.0 & $1.2 \times 10^{31}$ & $6.3 \times 10^{20}$ &      0.452594 & \citet{Traulsen20} \\
J004923.13+320036.5 & 1.586965 &       CHANDRA &     1218.0 & $1.7 \times 10^{30}$ & $5.2 \times 10^{20}$ &      0.112231 &    \citet{Evans10} \\
J005146.51-715953.5 & 0.870922 &    XMM-NEWTON &            &                      &                      &      0.606258 & \citet{Traulsen20} \\
\bottomrule
\end{tabular}

    \end{center}
\end{table*}

\section{Conclusions} \label{sec:conclusion}

We present a catalog of detached eclipsing binary parameters for \nSOL{} ASAS-SN eclipsing binaries. After refining the ASAS-SN $V$-band period from \cite{JayasingheII} with the $g$-band data, we use \PHOEBE{} to model the ASAS-SN data to determine the sum of fractional radii, ratio of effective temperatures, inclination, and eccentricity. We visually inspect all \nEA{} light curve solutions, and select \nSOL{} solutions for our final catalog. As a part of visual inspection, we use the TESS light curves from the SPOC \citep{Caldwell20} and QLP pipelines \citep{Huang20a, Huang20b, Kunimoto21} to validate the orbital period results. We identify a range of physically interesting features in the parameter distributions: 
\begin{itemize}
    \item The eccentricity distribution expands with increasing period, and we identify \nEccpV{} systems with $e>0.5$. High eccentricity systems are, in general, less detectable as eclipsing binaries than as spectroscopic binaries due to selection effects at longer periods. 
    \item The period distribution varies with absolute magnitude, reflecting the physical limits and detectability of stars of different radii in detached systems. We use the Gaia CMD to identify systems with subgiant and giant components. 
    \item The \teffratio{} and \requivsumfrac{} populations are both bimodal. The low \teffratio{} Group II component is associated with larger \requivsumfrac{}. The Group II systems tend to be more luminous and have light curves that begin to look more like semidetached binaries while the Group I systems are clearly detached systems. The Group II systems also show a deficit of edge on inclinations, which could be explained if the high inclination Group II systems were classified as semidetached or be a systematic effect from the fixed primary temperature. 
    \item While the individual fractional radii are not well constrained, there are two clear peaks in the distribution of the systems in the $\rho_1$-$\rho_2$ plane. This was noticed by \citet{Devor05} who suggested it was due to period confusion. Our visual inspection of the TESS light curves shows that this is not the case. 
\end{itemize}

Finally, we match the catalog of detached EBs with X-ray catalogs to identify \nXRAY{} systems with evidence of chromospheric or coronal activity that may be of interest for followup. We also identify \nFLAG{} systems with more physically complicated TESS light curves that will be the focus of a subsequent paper. The binaries included in this catalog span a wide range of stellar parameter space for periods ranging from a less than a day to more than 100 days. Interesting subsets of the binaries include short period eccentric systems and systems with subgiant and giant primaries that can easily be identified for detailed study. 

\section*{Acknowledgements}

We thank the {\tt PHOEBE} developers for their help in troubleshooting. KZS thanks Yanqin Wu for interesting discussions.

We thank Las Cumbres Observatory and its staff for their continued support of ASAS-SN. ASAS-SN is funded in part by the Gordon and Betty Moore Foundation through grants GBMF5490 and GBMF10501 to the Ohio State University, and also funded in part by the Alfred P. Sloan Foundation grant G-2021-14192.

DMR, TJ, KZS and CSK are supported by NSF grants AST-1814440 and AST-1908570. TJ acknowledges support 
from the Ohio State Presidential Fellowship. TAT is supported in part by NASA grant 80NSSC20K0531. B.J.S. is supported by NSF grants AST-1920392, AST-1911074, and AST-1911074.

This work has made use of data from the European Space Agency (ESA)
mission {\it Gaia} (\url{https://www.cosmos.esa.int/gaia}), processed by
the {\it Gaia} Data Processing and Analysis Consortium. 

This paper includes data collected with the \textit{TESS} mission, obtained from the MAST data archive at the Space Telescope Science Institute (STScI). Funding for the TESS mission is provided by the NASA Explorer Program. STScI is operated by the Association of Universities for Research in Astronomy, Inc., under NASA contract NAS 5-26555.CSK, KZS and DMR TESS research is supported by NASA grant 80NSSC22K0128.

\section*{Data Availability}

The ASAS-SN photometric data underlying this article are available in the ASAS-SN eclipsing
binaries database (https://asas-sn.osu.edu/binaries) and the ASAS-SN Photometry Database
(https://asas-sn.osu.edu/photometry). 
The data underlying this article are available in the article and in its online supplementary material.


\clearpage{}
\bibliographystyle{mnras}
\bibliography{asassn_ebs, heasarc_master} 


\appendix
\section{Selected Light Curve Fits} \label{appendix:LCs}

\begin{figure}
    \centering
    \includegraphics[width=\linewidth]{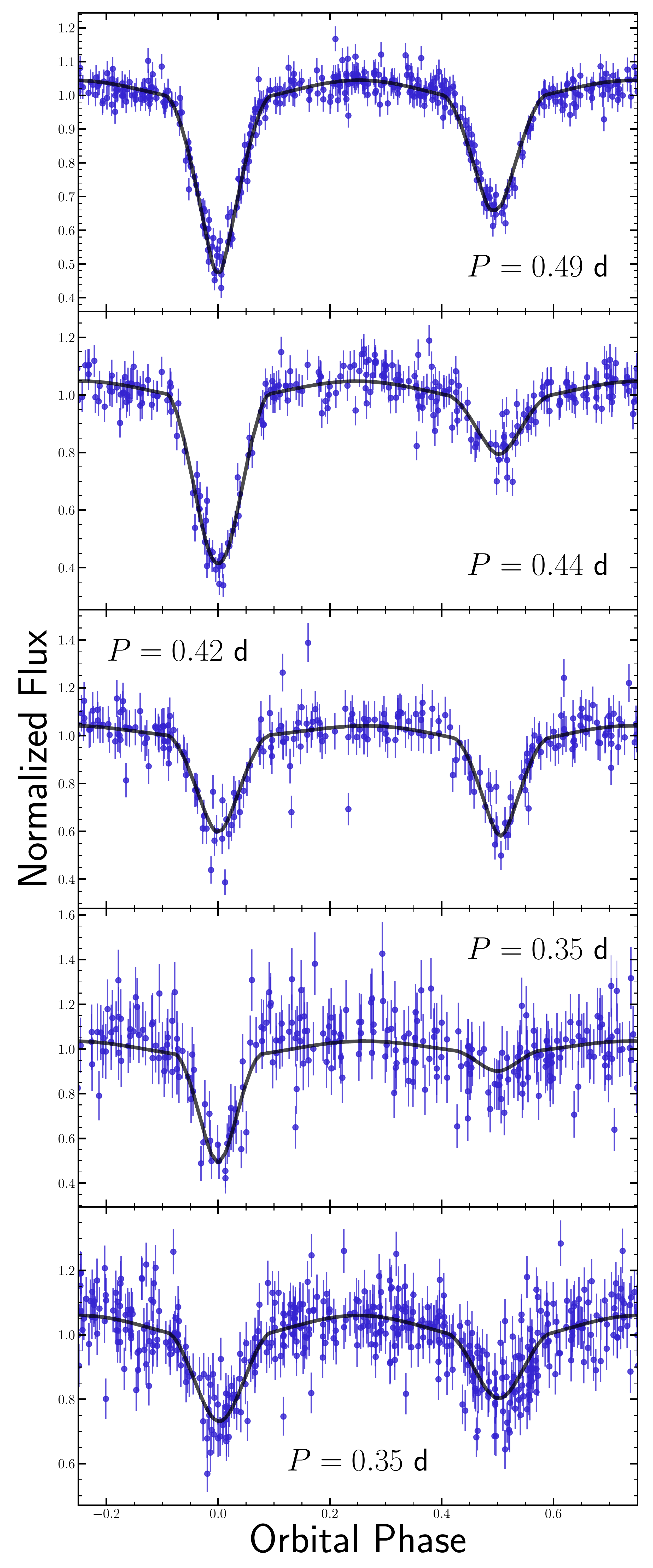}
    \caption{Examples of the shortest period binaries and their models. The phase range of $-0.25$ to $+0.75$ is used to clearly show both eclipses.}
    \label{fig:panel_short_period}
\end{figure}

\begin{figure}
    \centering
    \includegraphics[width=\linewidth]{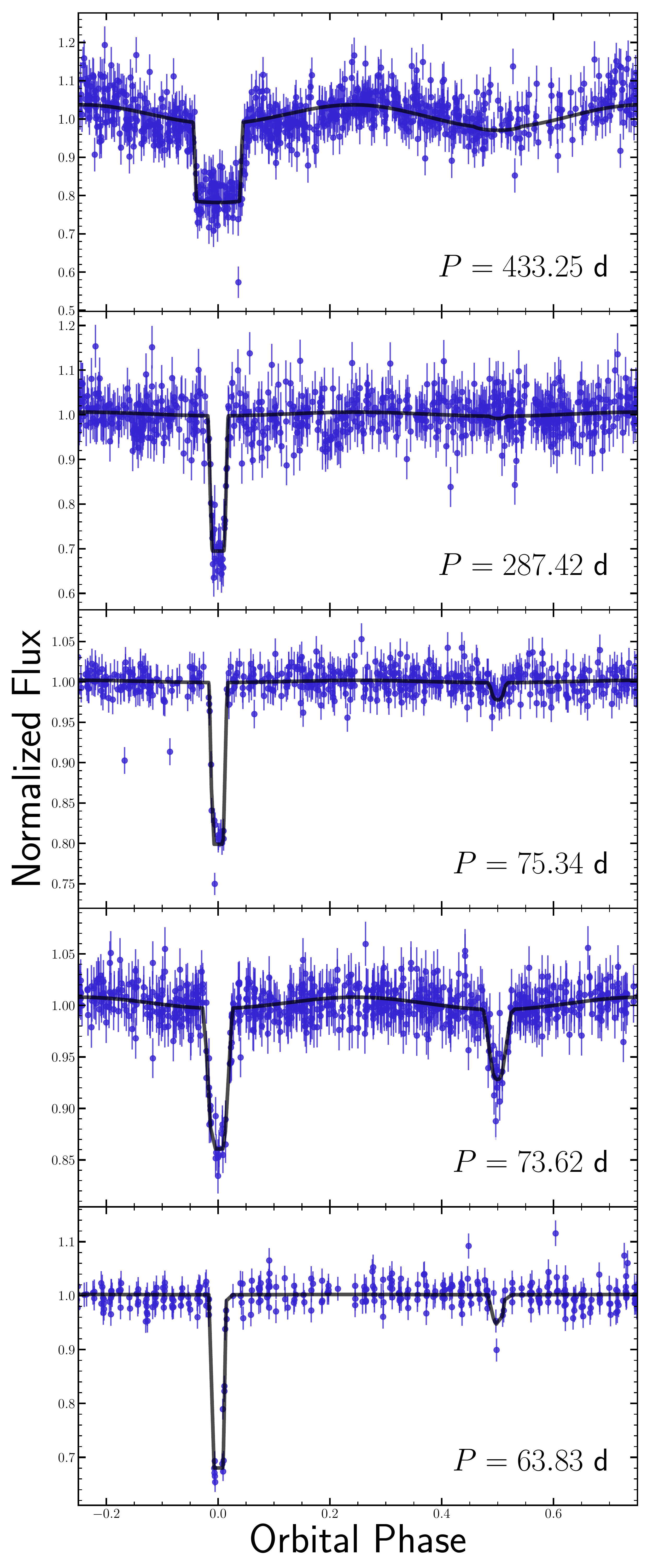}
    \caption{Same as Figure \ref{fig:panel_short_period}, but for some of the longest period variables.}
    \label{fig:panel_long_period}
\end{figure}

\begin{figure}
    \centering
    \includegraphics[width=\linewidth]{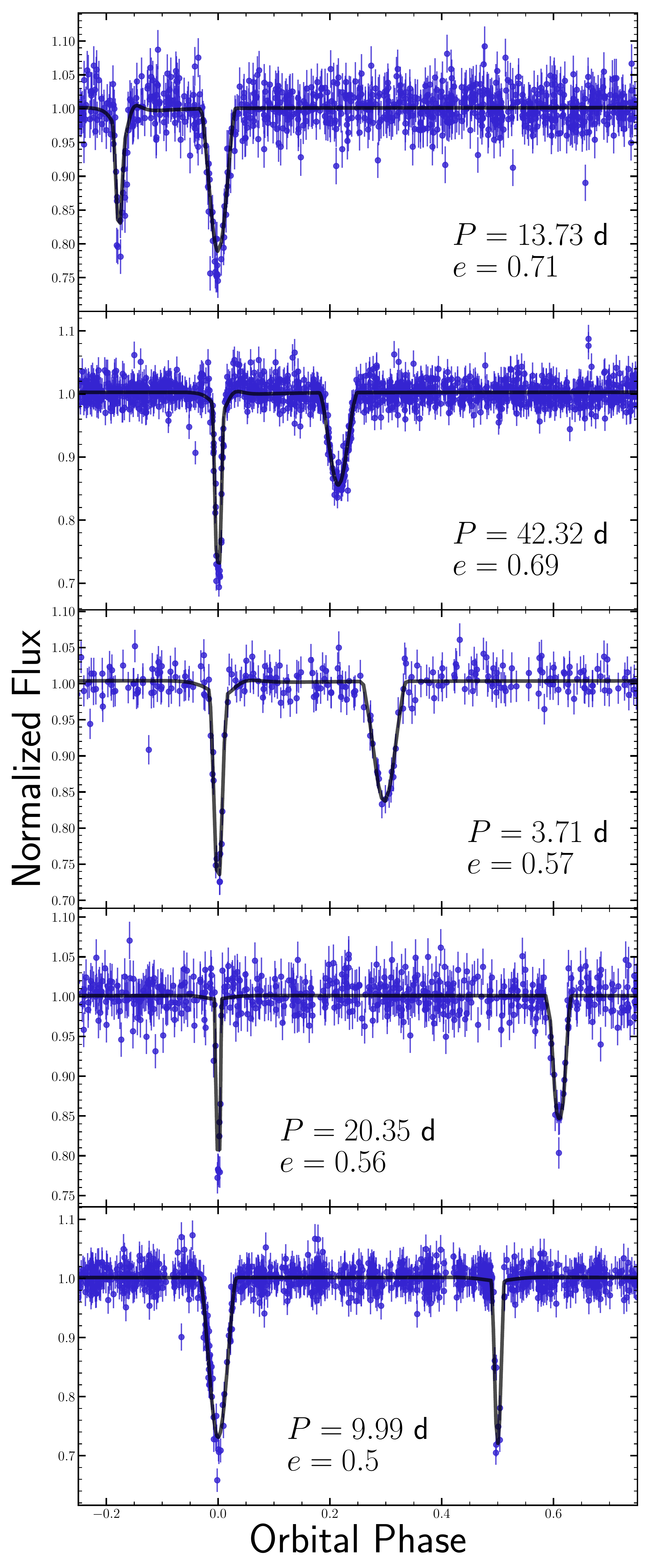}
    \caption{Same as Figure \ref{fig:panel_short_period}, but for high eccentricity systems.}
    \label{fig:panel_high_ecc}
\end{figure}

\begin{figure}
    \centering
    \includegraphics[width=\linewidth]{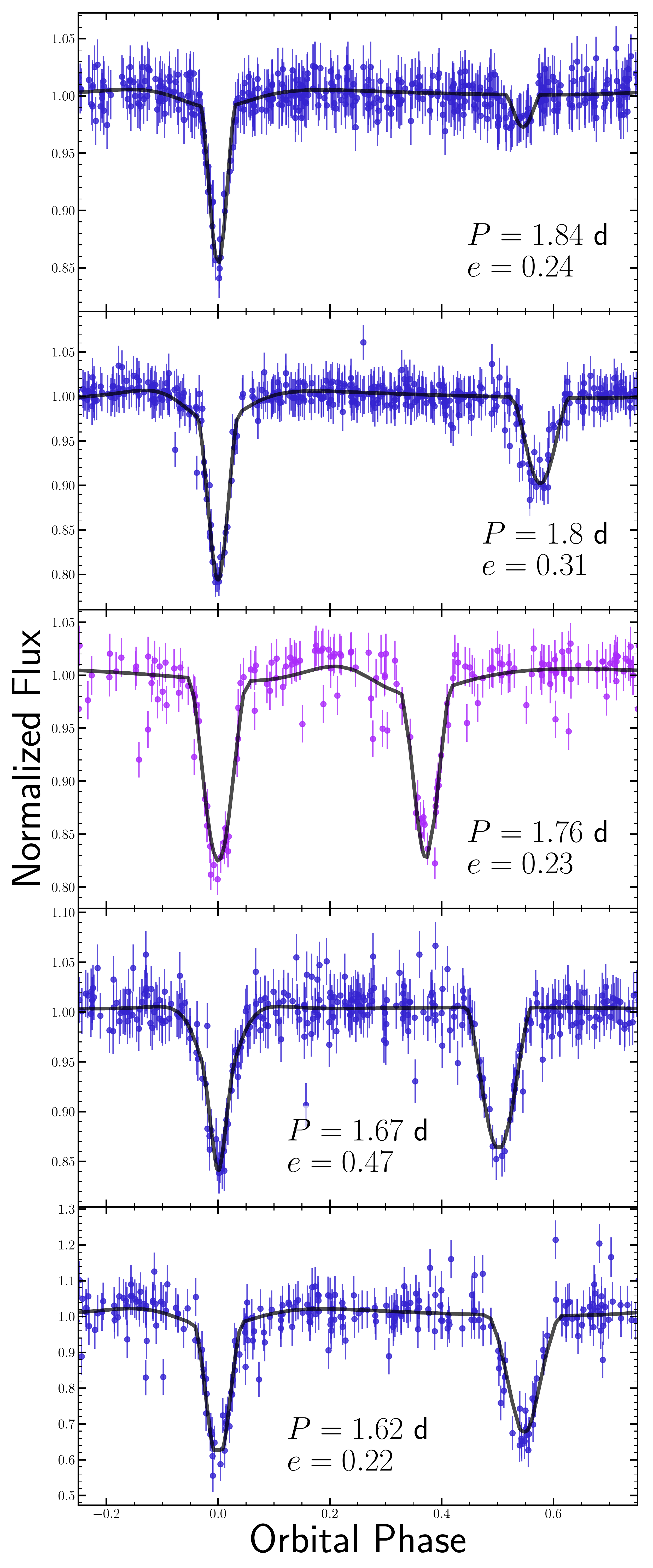}
    \caption{Same as Figure \ref{fig:panel_short_period} but for EBs with short periods and high eccentricities.}
    \label{fig:panel_short_p_high_ecc}
\end{figure}

\begin{figure}
    \centering
    \includegraphics[width=\linewidth]{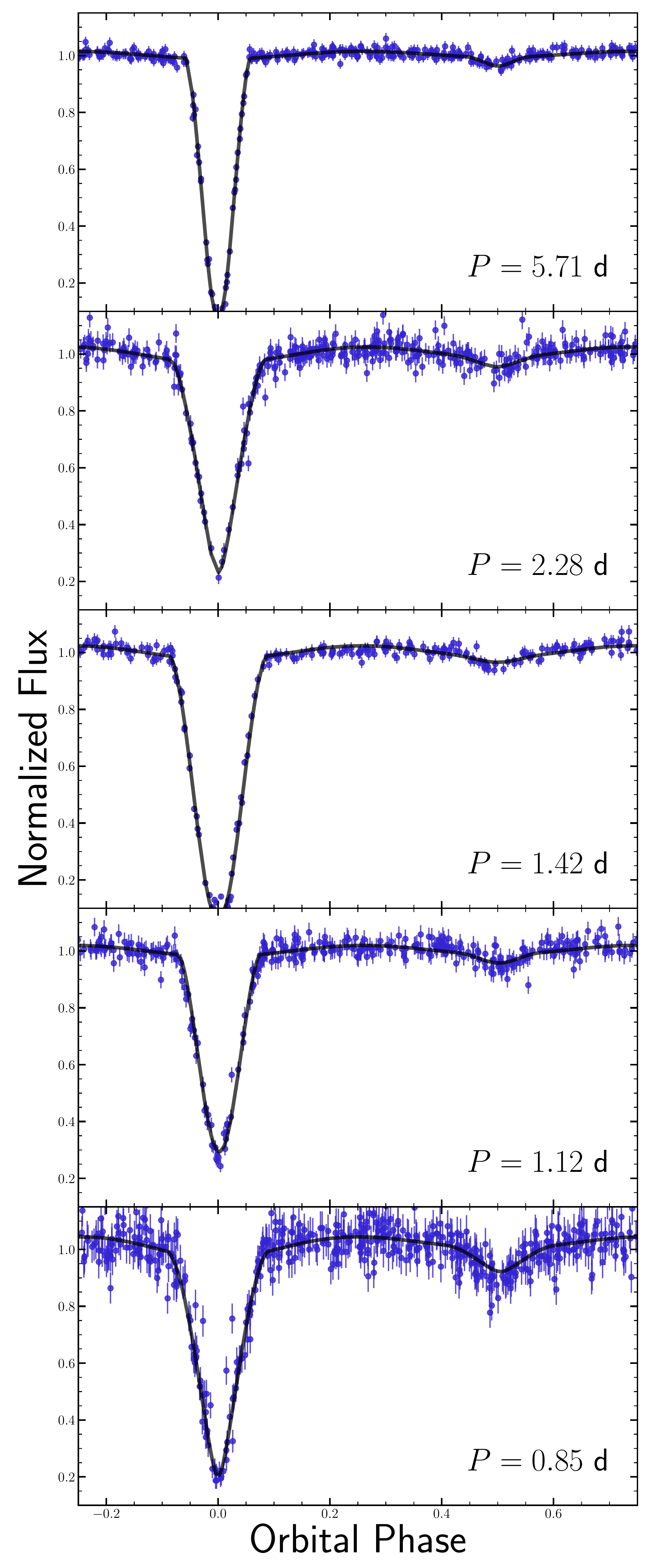}
    \caption{Same as Figure \ref{fig:panel_short_period} but for EBs with deep primary eclipses.}
    \label{fig:panel_deep_primary}
\end{figure}

\begin{figure}
    \centering
    \includegraphics[width=\linewidth]{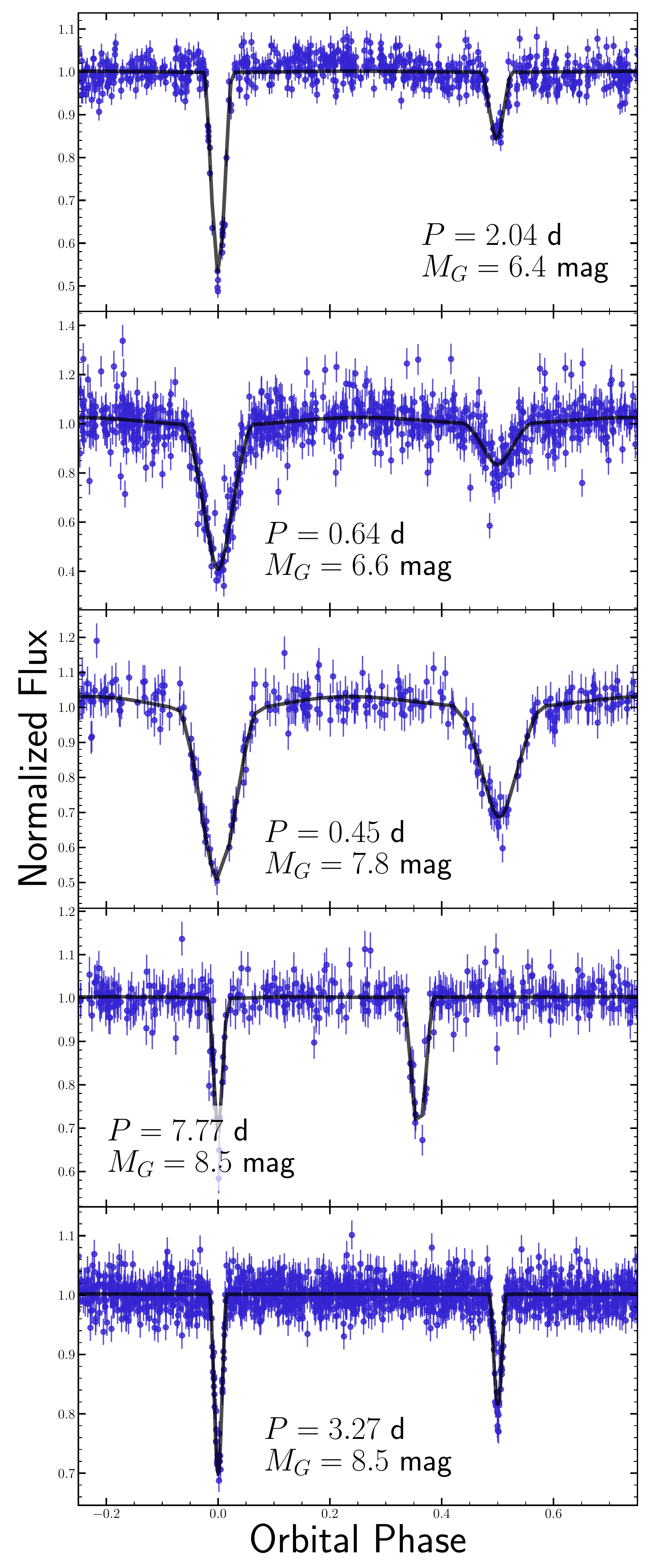}
    \caption{Same as Figure \ref{fig:panel_short_period} but for low-luminosity EBs. The absolute Gaia $G$-band magnitude corrected for extinction is given in the upper right of each panel.}
    \label{fig:panel_least_luminous}
\end{figure}

All light curve fits from in Table \ref{tab:ea_table} will be included in the ASAS-SN Variable Stars Database. Here we show subsets of light curves that may be of particular interest for followup. Figure \ref{fig:panel_short_period} shows some of the shortest period EBs in the catalog. Although some of these could be classified as semi-detached, we only removed obvious contact systems during visual inspection. Figure \ref{fig:panel_long_period} shows the longest period EBs. Figure \ref{fig:panel_high_ecc} shows systems with high eccentricitiy, and Figure \ref{fig:panel_short_p_high_ecc} shows short period systems with non-zero eccentricity. Figure \ref{fig:panel_deep_primary} shows systems with deep primary eclipses and Figure \ref{fig:panel_least_luminous} shows binaries with low luminosity $G$-band magnitudes. The light curves for these targets in Figures \ref{fig:panel_short_period}~to~\ref{fig:panel_least_luminous} are shown on a Gaia CMD in Figure \ref{fig:cmd_lcs}. 

\begin{figure}
    \centering
    \includegraphics[width=\linewidth]{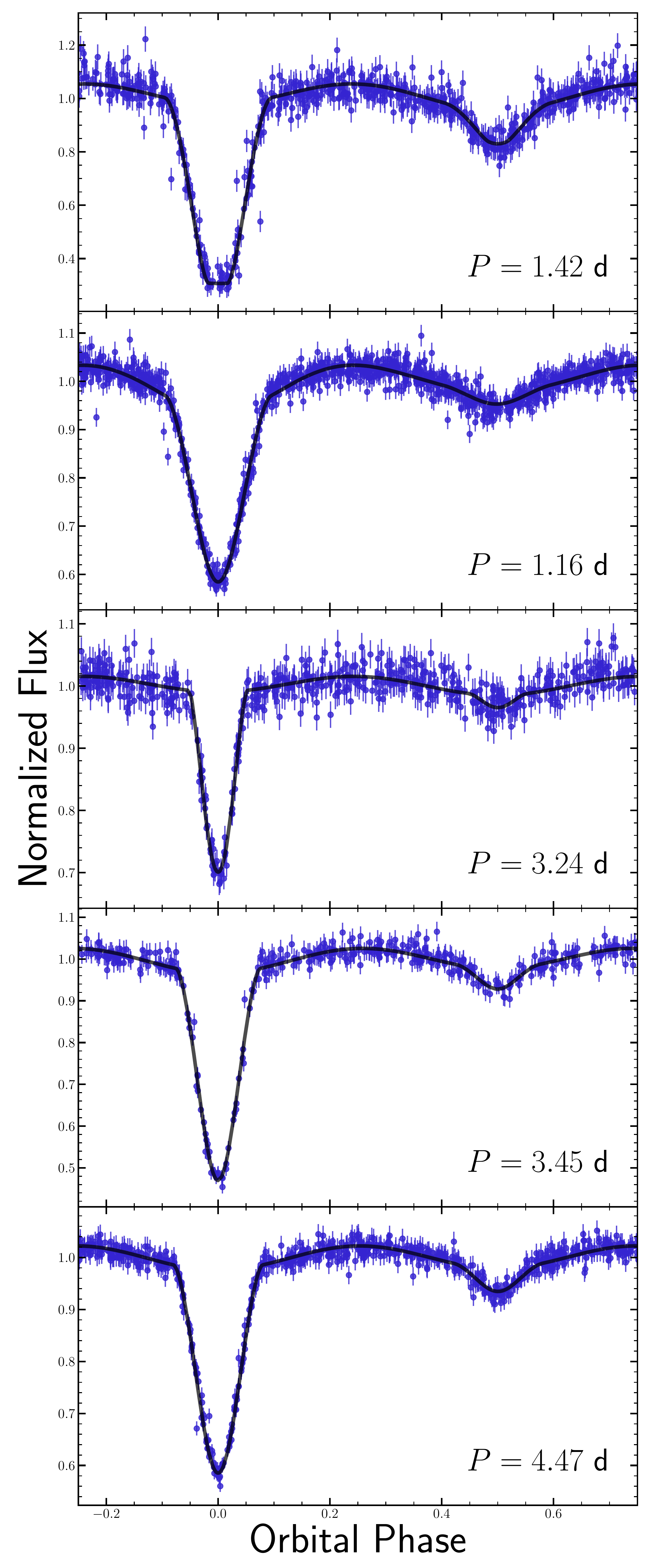}
    \caption{Same as Figure \ref{fig:panel_short_period} but for Group II EBs identified in Section \ref{sec:teffratio_and_radii}.}
    \label{fig:panel_groupII}
\end{figure}

In Section \ref{sec:teffratio_and_radii} we identified two groups in the \requivsumfrac{} and \teffratio{} parameter space. Figure \ref{fig:panel_groupII} shows examples for Group II systems that resembly semidetached binary light curves.

\begin{figure}
    \centering
    \includegraphics[width=\linewidth]{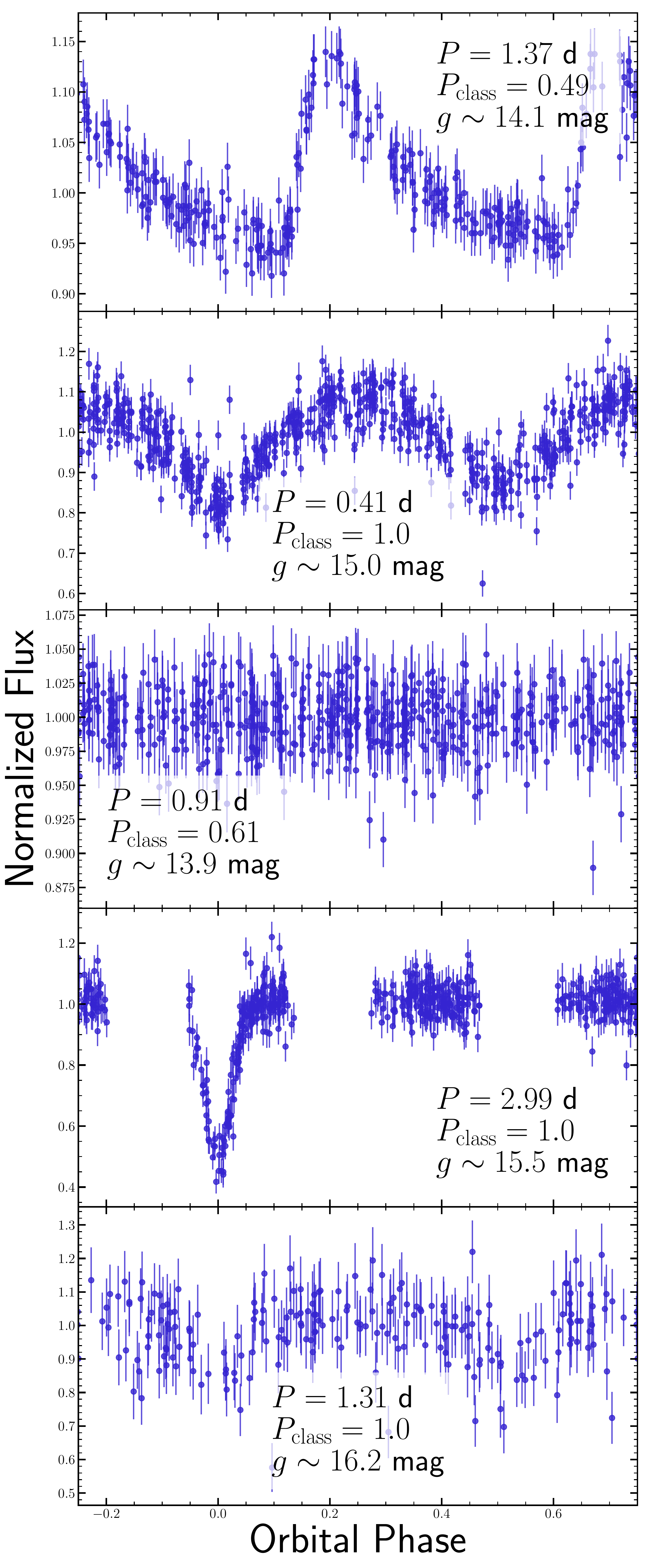}
    \caption{Examples of targets rejected during visual inspection.  From top to bottom, the light curves show an RR Lyrae, a contact binary, a non-variable target, a EB with a near 3 day period and poor phase sampling, and a faint target. The period, classification probability $P_{\rm{class}}$, and median $g$-band magnitude are labeled in each panel.}
    \label{fig:panel_poor_ea}
\end{figure}

During visual inspection we removed \nPOOR{} systems classified as EAs in \citet{Jayasinghe21} from the catalog. Figure \ref{fig:panel_poor_ea} shows examples of the types of systems removed as well as their classification probabilities, $P_{\rm{class}}$, from \citet{JayasingheII}.

\begin{figure*}
    \centering
    \includegraphics[width=\linewidth]{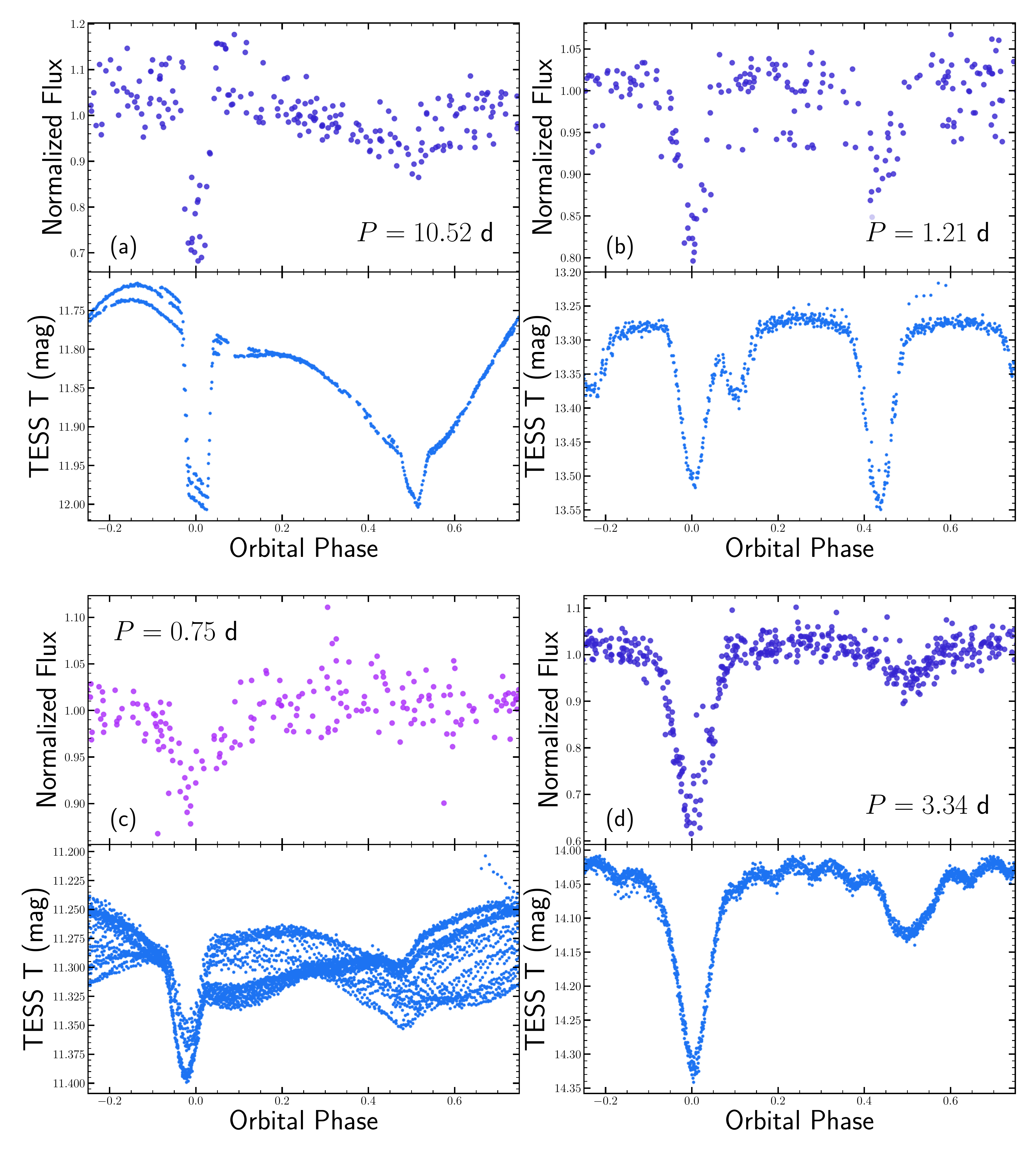}
    \caption{Examples of targets with extra physics identified during visual inspection. We show the ASAS-SN $g$-band and TESS $T$-band light curves from the QLP pipeline \citep{Huang20a, Huang20b, Kunimoto21}. Panel (a) may be an eclipsing cataclysmic variable \citep[e.g.,][]{Feline04}. The system in panel (b) is a quadruple system (V0849 Aur) with a near 3:2 period ratio \citep{Cagas12}. The system in panel (c) shows rapid changes in shape likely due to spot modulation. Finally, the system in panel (d) shows evidence of additional pulsations, but the ATLAS REFCAT 2 catalog \citep{Tonry18} suggests there are multiple nearby sources, so this could be an example of a blend.}
    \label{fig:panel_extra_physics}
\end{figure*}

\begin{figure*}
    \centering
    \includegraphics[width=\linewidth,trim={0 0 2cm 0}]{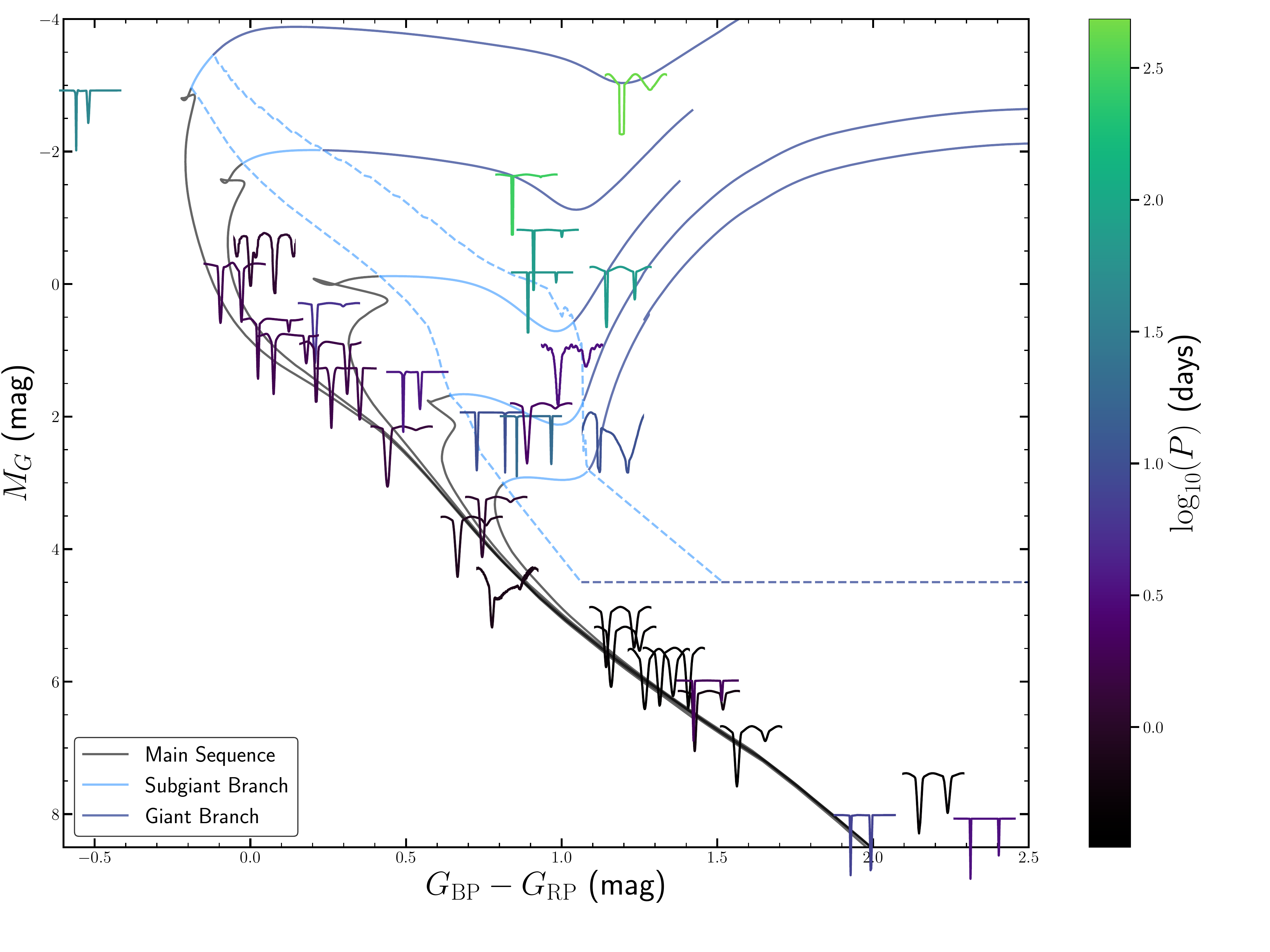}
    \caption{\PHOEBE{} model fits for the systems shown in Figures \crefrange{fig:panel_short_period}{fig:panel_least_luminous} and \ref{fig:panel_extra_physics} on a Gaia CMD colored by $\log P$. For the extra physics targets from Figure \ref{fig:panel_extra_physics}, the median filtered TESS light curve is shown instead of the \PHOEBE{} fit.}
    \label{fig:cmd_lcs}
\end{figure*}

Finally, during visual inspection we identified \nFLAG{} systems as potentially having spots, evidence of mass transfer, or potential third bodies. Many of these systems were identified through visual inspection of the TESS light curves. Figure \ref{fig:panel_extra_physics} shows the ASAS-SN $g$-band and TESS $T$-band light curves four examples of systems with more complex light curves.

\bsp	
\label{lastpage}
\end{document}